\documentclass[conference]{IEEEtran}

\usepackage[style=ieee,backend=biber]{biblatex}
\AtBeginBibliography{\footnotesize} 
\usepackage[framed,autolinebreaks,useliterate]{mcode}
\usepackage{mathtools}
\usepackage{nicematrix}
\usepackage{float}
\usepackage{tabularx}
\usepackage{booktabs}
\usepackage{enumitem} 

\usepackage{graphicx} 
\graphicspath{ {images/} }

\usepackage{color, colortbl}
\usepackage{multirow}

\usepackage[linesnumbered,ruled,vlined]{algorithm2e} 

\usepackage[utf8]{inputenc}
\usepackage[english]{babel}
\usepackage{amsthm,amssymb}
\usepackage{amsmath}
\usepackage{bm}
\usepackage{optidef}
\usepackage{mathrsfs}

\usepackage[font=small,labelfont=bf,labelsep=space]{caption}
\usepackage{xcolor}
\def\BibTeX{{\rm B\kern-.05em{\sc i\kern-.025em b}\kern-.08em
    T\kern-.1667em\lower.7ex\hbox{E}\kern-.125emX}}

\usepackage{tikz}
\usetikzlibrary{matrix,backgrounds,positioning}
\theoremstyle{definition}
\newtheorem{example}{Example}
\newtheorem{theorem}{Theorem}
\newtheorem{proposition}{Proposition}
\newtheorem{lemma}{Lemma}
\newtheorem{corollary}{Corollary}

\newtheorem{definition}{Definition}

\newtheorem{remark}{Remark}

\DeclareMathOperator{\supp}{supp}

\DeclareMathOperator{\w}{w}
\DeclareMathOperator{\dist}{d}

\newcommand{\wm}{\w_{\min}}

\newcommand{\A}{\mathcal{A}}

\newcommand{\I}{\mathcal{I}}

\renewcommand{\H}{\mathcal{H}}
\renewcommand{\S}{\mathcal{S}}

\newcommand{\T}{\mathcal{T}}

\newcommand{\J}{\mathcal{J}}

\newcommand{\M}{\mathcal{M}}

\newcommand{\C}{\mathcal{C}}
\newcommand{\Ci}{\mathcal{C}_i}
\newcommand{\CI}{\mathcal{C}(\mathcal{I})}
\newcommand{\CiI}{\mathcal{C}_i(\mathcal{I})}
\newcommand{\G}{\mathcal{G}}

\newcommand{\K}{\mathcal{K}}

\newcommand{\Amr}{\A_{m-r}}
\newcommand{\bA}{\mathbf{A}}
\newcommand{\bB}{\mathbf{B}}
\newcommand{\bgi}{\mathbf{g}_i}
\newcommand{\bgj}{\mathbf{g}_j}
\newcommand{\bgm}{\mathbf{g}_m}

\newcommand{\bgh}{\mathbf{g}_h}
\newcommand{\bg}{\mathbf{g}}

\newcommand{\barf}{\bar{f}}
\newcommand{\barg}{\bar{g}}

\newcommand{\bve}{\bm{\varepsilon}}
\newcommand{\bvg}{\bm{\gamma}}

\newcommand{\bGN}{\boldsymbol{G}_N}
\newcommand{\bc}{\boldsymbol{c}}

\newcommand{\bx}{\boldsymbol{x}}

\newcommand{\bv}{\boldsymbol{v}}

\newcommand{\bG}{\boldsymbol{G}}

\newcommand{\ind}{\operatorname{ind}}
\newcommand{\ev}{\operatorname{ev}}

\newcommand{\bi}{\bm{i}}
\newcommand{\ff}{\mathbb{F}}
\newcommand{\ft}{\mathbb{F}_2}

\newcommand{\Alow}{{\rm LTA}(m,2)}

\newcommand{\GL}{{\rm GL}(m,2)}
\newcommand{\Mon}{\mathcal{M}_{m}}
\newcommand{\weako}{\preceq_w}
\newcommand{\Rm}{{\mathbf {R}}_m}

\IEEEoverridecommandlockouts                              

\title{
On the Closed-form Weight Enumeration\\ of Polar~Codes: 1.5$d$-weight codewords
}

\author{
\IEEEauthorblockN{Mohammad Rowshan$^{\ast\dagger}$, Vlad-Florin Drăgoi$^{\ast\ddagger}$, and Jinhong Yuan$^\dagger$, {\em Fellow, IEEE}}
\IEEEauthorblockA{$^\dagger$School of Electrical Eng. and Telecom., University of New South Wales (UNSW), Sydney, Australia\\
$^\ddagger$Faculty of Exact Sciences, Aurel Vlaicu University, Arad, Romania\\
m.rowshan@unsw.edu.au, vlad.dragoi@uav.ro, j.yuan@unsw.edu.au }
\thanks{$^\ast$These authors contributed equally (Corresponding author: Vlad-Florin Drăgoi).}
}

\begin{document}

\maketitle
\pagestyle{plain}

\begin{abstract}
The weight distribution of error correction codes is a critical determinant of their error-correcting performance, making enumeration of utmost importance. In the case of polar codes, the minimum weight $\wm$ (which is equal to minimum distance $d$) is the only weight for which an explicit enumerator formula is currently available. Having closed-form weight enumerators for polar codewords with weights greater than the minimum weight not only simplifies the enumeration process but also provides valuable insights towards constructing better polar-like codes. 
In this paper, we contribute towards understanding the algebraic structure underlying higher weights by analyzing Minkowski sums of orbits. Our approach builds upon the lower triangular affine (LTA) group of decreasing monomial codes. Specifically, we propose a closed-form expression for the enumeration of codewords with weight $1.5\wm$. Our simulations demonstrate the potential for extending this method to higher weights. 
\end{abstract}


\section{Introduction}\label{sec:intro}
Polar codes \cite{arikan} are the first class of channel codes with explicit construction that achieve the symmetric (Shannon) capacity of a binary-input discrete memoryless channel (BI-DMC). As short and medium-length polar codes have shown a remarkable error correction performance, they had been chosen as a coding scheme for logical control channels in the 5th generation of mobile broadband standard \cite{3GPP}. In the past decade, polar codes were the focus of attention in the field of coding theory. The efforts were mostly focused on low-complexity methods for code construction, and performance improvement through code concatenation and decoding schemes. The more fundamental problems such as the algebraic characteristics of polar codes were less investigated. One of the major problems is the algebraic investigation of weight distribution and finding closed-form expressions for the enumeration of the codewords with certain weights. 

The weight distribution of a code determines the error correction performance of a code. The upper bound for the block error rate (BLER) of linear codes under maximum likelihood (ML) decoding can be obtained by Union bound 
where the number of codewords with small weights, such as the minimum weight, the second minimum weight, and larger ones take the role of coefficient in the largest terms \cite[Sect. 10.1]{lin_costello}. 
It was shown in \cite{bardet2016crypt} that polar codes along with Reed-Muller codes belong to a larger family of codes named decreasing monomial codes. On top of the polynomial structure that both Reed-Muller and polar codes possess, the decreasing property induces new algebraic properties. Exploited for the first time in \cite{bardet2016crypt,bardet} they allowed the discovery of new structural properties such as duals of decreasing monomial codes are still decreasing monomial codes, the permutation group of decreasing monomial codes admits as a subgroup, the lower triangular affine group ($\Alow$). Having a slightly better understanding of the algebraic structure of polar codes allowed the scientific community to propose practical applications. In \cite{bardet} the permutation group revealed the complete structure of the minimum weight codewords of decreasing monomial codes, thus implicitly of polar codes. In \cite{mondelli-construction} a sub-linearity code construction was proposed. The permutation group has a significant contribution in parallelized decoding of polar codes \cite{ZWT21, GEECB21, GEEB21, PBL21, PBL22, IU22,IU22a}. Also, the permutation group was used as an optimisation tool in the weight enumeration algorithm of Yao, Fazeli and Vardy \cite{yao}.

Finding closed-form expressions for the weight distribution of Reed–Muller codes is still an open problem after more than five decades since notable progress was made towards this goal. Kasami and Tokura \cite{kasami1970weight} characterized the codewords of Reed–Muller codes of weight up to twice the minimum weight of the code. The results were extended to the enumeration of codewords with weights less the 2.5 times the minimum weight in \cite{kasami1976w2.5d}. In the case of polar codes, the first progress was made by the closed-form enumeration of minimum weight codewords in \cite{bardet} and it was stopped there. 

Due to the importance of the weight distribution of polar codes, the rest of the attempts were focused on the algorithmic solutions for either exact or approximate enumeration of polar codewords with various weights. 
In \cite{liu_analys}, the authors proposed to send the all-zero codeword over a channel with low noise or to receive at very high SNRs, and count the re-encoded messages with certain weights at the output of a successive cancellation list decoder with very large list size. The method presented in \cite{valipour} suggests  computing a probabilistic weight distribution expression efficiently. The authors in \cite{zhang_prob} proposed a way to get an approximate distance spectrum of polar codes with large lengths using the spectrum of short codes and a probabilistic assumption on appearing ones in codewords. Based on the weight distribution of $|u|u+v|$ constructed codes in \cite{fossorier}, the weight distribution of the words generated by polar transform was found recursively in \cite{polyanskaya}. This work and the work in \cite{bardet} inspired the authors of \cite{yao} to propose a deterministic recursive algorithm to count all the codewords of polar codes with any weights although due to high complexity, it cannot be used for medium and long block-lengths. 

In this work, we take one step forward and partially address the long-standing problem of weight distribution for polar codes by enumerating codewords of weight $1.5\wm.$ Our result applies to any decreasing monomial code.  In order to address this problem we need to understand that closed formulae for enumeration of codewords of a given weight are indeed complex and challenging problems for polar codes. Dealing with the general case of decreasing monomial codes has several advantages and could unify old results (the case of Reed-Muller codes) with the case of polar codes. In order to achieve our goal we built our results on two key ingredients which are the permutation group ($\Alow$) and a classification theorem by Kasami and Tokura for weight smaller than $2\wm$ \cite{kasami1970weight}. The theorem in \cite{kasami1970weight} applies to any polynomial code and gives the polynomial shape (up to linear affine transformation) of any codeword of weight smaller than $2\wm.$ Such codewords can be defined as sums of minimum weight codewords, which are defined as the evaluation of polynomials in orbits $\Alow\cdot f$ for a monomial $f$ of maximum degree, let's say $\deg(f)=r$. Hence, the weight of such a sum is given by the number of positions on which two minimum weight codewords overlap. This leads us to one of the key ingredients in our quest, i.e., the understanding of the Minkowski sum of two orbits $\Alow\cdot f+\Alow \cdot g$ where $f,g$ are monomials of maximum degree ($f,g$ define minimum weight codewords). 

The initial step involves transposing the findings from \cite{kasami1970weight} to the scenario of decreasing monomial codes. To accomplish this, we will establish the following:
\begin{itemize}
    \item find the subgroups that generate the complete orbits for sums and product of sums;
    \item given a codeword $\bc$ of weight $1.5\wm$, compute the maximum degree monomials, say $f$ and $g$ s.t. $\bc$ is the evaluation of a polynomial that belongs to the set $\Alow\cdot f+\Alow\cdot g.$  
\end{itemize}

The next step will be to determine the cardinal of a Minkowski sum of orbits. We demonstrate that if there is a particular order relation on the variables of $f$ and $g$ then, the cardinal of $\Alow\cdot f+\Alow\cdot g$ is equal to the product of cardinals of the two sets $\Alow\cdot f$ and $\Alow\cdot g.$
To prove our result we will introduce the concept of polynomial collision. Informally, two pair of distinct polynomials $(P,Q),(P^{*},Q^{*})\in \Alow\cdot f\times \Alow\cdot g$ form a collision if $P+q=P^{*}+Q^{*}.$ We demonstrate that the existence of a collision at the level of monomials $(f,g)$, by a factorization procedure (which is possible due to the decreasing order relation), can be transposed down to $(f/h,g/h)$ where $h=\gcd(f,g).$ Next, we characterize all possible cases of collisions and thus retrieve the set of all invariants. A direct consequence of our result is that we give a simple formula to compute the cardinal of a Minkowski sum $\Alow\cdot f+\Alow\cdot g$ for $f$ and $g$ of degree 2. 

While the aforementioned results are rather theoretical, we do apply them for a practical challenge. We determine closed formulae for weight $1.5\wm.$ Let's start by recalling that words of weight $1.5\wm$ belong to sum of orbits of some monomials $f$ and $g$ of maximum degree $\deg(f)=\deg(g)=r$ and such that $\deg(\gcd(f,g))=r-2.$ We demonstrate that distinct pairs of such monomials define disjoint Minkowski sums (property analogue to disjoint orbits for minimum weight codewords). Finally, we give a formula for counting such codewords and test it for different polar codes. We retrieve well-known results for small-length codes as well as the formulas for Reed-Muller codes. We push a bit further our simulations and provide weight counting for large polar codes.

\section{Preliminaries} 
\label{sec:pre}
 \subsection{Basic Concepts in Coding Theory and Notations}
 \label{subsec:basic}

We denote by $\ft$ the finite field with two elements and by $\oplus$ the addition operator in this field. 
Also, subsets of consecutive integers are denoted by $[\ell,u]\triangleq\{\ell,\ell+1,\ldots,u\}$, and by  $[n]\triangleq[0,n-1].$
The \emph{support} of a vector $\bc = [c_0,\ldots,c_{N-1}] \in \ft^N$ is the set of indices where $\bc$
has a nonzero coordinate, i.e. $\supp(\bc) \triangleq \{i \in [N] \mid c_i \neq 0\}$. 
The cardinality of a set is denoted by $|\cdot|$ and the set difference by $\backslash$. 
The Hamming \emph{weight} of a vector $\bc \in \ft^N$ is $\w(\bc)\triangleq |\supp(\bc)|$. 
Given two vectors $\bc = [c_0, c_1, \ldots, c_{N-1}]$ and $\bc' = [c'_0, c'_1, \ldots, c'_{N-1}]$, 
the Hamming \emph{distance} between $\bc$ and $\bc'$ is defined to be the number of coordinates 
where $\bc$ and $\bc'$  deffer, namely, $\dist(\bc,\bc') = |\{i \in [N] \mid c_i \ne c'_i\}|$. 

A $K$-dimensional subspace $\C$ of $\ft^N$ is called a linear $(N,K,d)$ \emph{code} over $\ft$ 
if the minimum distance of $\C$, 
\[
d_{min} = \dist(\C) \triangleq \min_{\bc,\bc' \in \C, \bc \neq \bc'} \dist(\bc,\bc')=d.
\]
It is easy to see that the Hamming norm induces the Hamming distance and vice-versa. Hence, we have (see \cite[Section 3.3]{lin_costello})
 \[\wm\triangleq\min_{\bc\in\C,\bc\neq 0}\w(\bc)=\dist(\C).\] 
We usually use the short notation $(N,K)$ for codes where we refer to $N$ and $K$ as the \emph{length} and the \emph{dimension} of the code. The vectors in $\C$ are called \emph{codewords}.  
We also collect all codewords of code $\C$ with weight $\w$ in set $W_{\w}$ as
\[
    W_{\w}(\C) = \{ \bc\in\C \mid \w(\bc)=\w \}.
\]
A \emph{generator matrix} $\bG$ of an $(N,K)$ code $\C$ is a $K \times N$ matrix in $\ft^{K\times N}$ whose rows are $\ft$-linearly independent codewords of $\C$. Then $\C = \{\bv \bG \colon \bv \in \ft^K\}$.

Under a binary input additive white Gaussian noise (BI-AWGN) channel at a high signal-to-noise ratio (SNR) per information bit $E_b/N_0$, according to \cite[Sect. 10.1]{lin_costello}, 
the block error rate (BLER) of linear codes under soft-decision maximum likelihood (ML) decoding is upper bounded by 
\begin{equation}\label{eq:union_bound}
    \begin{split}
    P_e^{ML} & \leq \sum_{\w=\wm}^{N} A_{\w} Q(\sqrt{2\w\cdot R \cdot E_b/N_0}), 
    \end{split}
\end{equation} 
where $A_{\w}$ denotes the number of $\w$-weight codewords (or equivalently $A_{\w}=|W_{\w}(\C)|$), $Q(\cdot)$ is the tail probability of the normal distribution $\mathcal{N}(0,1)$ to find the pairwise error probability for a transmitted sequence and its corresponding distorted received one, and $R\triangleq K/N$ is the code rate. In the literature, $A_{\wm}$ is known as \emph{error coefficient} since it is the coefficient of the largest term for calculating the BLER upper bound. One can observe that a code with smaller $A_{\wm}$ is expected to provide a smaller BLER than a code with larger $A_{\wm}$, assuming they both have identical $\wm$. Note that this paper considers other dominant terms to give a more accurate measure of a code by finding $A_{\w}$ for all $\w<2\wm$, in particular for $\w=1.5\wm$.

\subsection{Monomial Codes}\label{ssec:monmial_codes}
A monomial is a single-term algebraic expression indicating the product of any subset of variables in $\mathbf{x}\triangleq(x_{0},\dots,x_{m-1})$. Assuming $\mathbf{i}=(i_{0},\dots,i_{m-1})$ where $i_j\in\{0,1\}$, an $m$-variate monomial is denoted as
$$\mathbf{x}^{\mathbf{i}}=\prod_{j=0}^{m-1}x_{j}^{i_{j}}=x_{0}^{i_{0}} \cdots x_{m-1}^{i_{m-1}}.$$
For simplicity, we denote a monomial by $f=\mathbf{x}^\mathbf{i}$. Moreover, let us denote the degree of a monomial $\mathbf{x}^{\mathbf{i}}$ as $\deg(\mathbf{x}^{\mathbf{i}})$(we have $\deg(\mathbf{x}^{\mathbf{i}})=\w(\mathbf{i})$)
and the set of all monomials by
$$
\Mon \triangleq \left\{\mathbf{x}^{\mathbf{i}} \mid\mathbf{i} \in \mathbb{F}_{2}^{m}\right\}.
$$
For any monomial $f\in\Mon$ of degree $1\leq s\leq m$ denoted as $f=x_{l_1}\dots x_{l_s}$ where $0\leq l_1 \leq l_2\dots \leq l_s\leq m-1$, the \emph{support of monomial} is denoted as $\ind(f)=\{l_1,\dots,l_s\}$. Observe that the support of a monomial $f=x_0^{f_0}\dots x_{m-1}^{f_{m-1}}$ includes all $j$ where $f_j=1$. The degree induces a ranking on any monomial set $\I\subseteq\Mon$, i.e., $\I=\bigcup_{j=0}^{m}\I_j$, where $\I_j=\{f\in\I\mid\deg(f)=j\}.$ 

Since we are interested in evaluations of monomials over entries in $\ft^m$, we 
will identify $x_i$ with $x_i^2$ and work in the ring $\Rm=\ft[x_0,\dots{},x_{m-1}]/(x_0^2-x_0,\dots,x_{m-1}^2-x_{m-1})$. The main reason of considering the aforementioned ideal is because we do not want to carry useless powers when multiplying polynomials or monomials. It will all become much clearer after explaining the connection between polynomials and codewords. 
Now, any subset $\I \subseteq \M_{m}$ forms a \emph{generating set} for a ($n=2^m, k=|\I|$) monomial code $\C$. 
This monomial code $\C$ as a linear code is the vector subspace $\CI \subseteq \mathbb{F}_{2}^{n}$ generated by the span of the row vectors as \emph{basis} resulting from the evaluation of the monomials in $\I$ \cite{bardet}. That is,
 $$\CI=\operatorname{span}(\{\ev(f) \mid f \in \I\}),$$
where $\ev(f)$ is the binary vector obtained by evaluating of monomial $f$ over all the binary entries in $\ft^m$, i.e., 
\[\ev(f)\triangleq (\ev(f)(0,\dots,0),\dots,\ev(f)(1,\dots,1)),\] where the set $\ft^m$ is ordered in decreasing order (see Example \ref{ex:1} for $m=2$). More exactly, order the entries $\bi=(i_0,\cdots,i_{m-1})\in\ft^m$ s.t. the left-most entry is the least significant bit (as the index of entries indicates) in the binary representation of $\bi$ and sort them in decreasing order. Since $\ev(f)$ is a binary vector it's Hamming weight is $\w(\ev(f))$ or simply $|f|_m$ and sometimes called the weight of $f$. We do have for any $f\in\Mon$, $\w(\ev(f))=2^{m-\deg(f)}.$ Indeed, the points of evaluation where $\ev(f)=1$ are defined by the subset $\{(f_0,\dots,f_{m-1})\in \ff_2^m\mid \forall i \in\ind(f) f_i=1\}.$ This set has cardinality $2^{m-|\ind(f)|}=2^{m-\deg(f)}$, since there are $m-\deg(f)$ free indices from $\ff_2.$   
\begin{example}\label{ex:1}
 Let $m=2$. The evaluation $\ev(f)$ of all monomials $f\in\M_n=\{\bm{1},x_0,x_1,x_0x_1\}$ and their corresponding rows in $\bG_N$ is as follows:
{
\begin{center}
\begin{tabular}{c|ccccc} 
  & $(i_0\;i_1):$ & 11 & 01 & 10 & 00 \\
\hline \hline 
 $\bg_0$ & $\mathrm{ev}\left(x_0 x_1\right)$ & 1 & 0 & 0 & 0 \\
 $\bg_1$ & $\mathrm{ev}\left(x_1\right)$ & 1 & 1 & 0 & 0 \\
 $\bg_2$ & $\mathrm{ev}\left(x_0\right)$ & 1 & 0 & 1 & 0 \\
 $\bg_3$ & $\operatorname{ev}(1)$ & 1 & 1 & 1 & 1
\end{tabular}
\end{center}
}
\noindent where, for instance, evaluation of $f=x_0$ gives 1 for all $(1,i_2), i_2\in\{0,1\}$ and 0 otherwise.
\end{example}

Since the function $\ev$ defines a vector space isomorphism between $\Rm$ and $\ft^n$ (see Corollary 3,2,6 in \cite{dragoi17thesis}) then given a monomial code $\CI$ with generator matrix $\mathbf{G}=(\bg_0 \cdots \bg_{k})^T$, for every row $\bg_i$ where $i\in[0,k)$ it exists a monomial $f\in\I$ such that $\ev(f)=\bg_i$. 
The minimum distance of a monomial code is then \cite{bardet}
\[
\min_{\mathbf{\bc} \in \CI} \w(\bc)=\min _{f \in \I} \w(\ev(f))=2^{m-r^+(\C)},
\]
where $r^+(\C)=\max _{f \in \I} \deg(f)$.

With the formalism defined above, the Reed-Muller code $\mathcal{R}(r, m)$ is a monomial code
$$
\mathcal{R}(r, m) \triangleq \operatorname{span}(\left\{\operatorname{ev}(f) \mid f \in \M_{m}, \deg(f)\leqslant r\right\}).
$$

We will also require to define the concept of sums/products of polynomial sets. 
\begin{definition}Given two polynomial sets $\S,\T\in \Rm$ their \emph{Minkowski sum} is $\S+\T=\{s+t\mid s\in \S,t\in \T\}$. Also, the product is defined as $\S\cdot\T=\{s\cdot t\mid s\in \S,t\in \T\}$, where $+$ and $\cdot$ stand for the polynomial addition and multiplication.
\end{definition}
\section{Decreasing monomial codes: the algebraic formalism behind polar codes}
\subsection{Polar Codes}
\label{subsec:RMPolar}
Polar codes of length $N=2^n$ are constructed based on the $n$-th Kronecker power, denoted by $(.)^{\otimes n}$, of binary Walsh-Hadamard matrix  
$\mathbf{G}_2 = 
{\footnotesize \begin{bmatrix}
1 & 0 \\
1 & 1
\end{bmatrix} }$, that is, $\bGN=\mathbf{G}_2^{\otimes n}$ which we call it {\em polar transform} throughout this paper. We denote polar transform by rows $\bg_i,i=[N]$ as $\bGN=[\bg_0,\bg_1,\cdots,\bg_{N-1}]^T$ where operator $T$ in $[\cdot]^T$ denotes the transpose of the matrix. 
The generator matrix of a polar code is formed by selecting a set of rows of $\bGN$. We use $\A$ to denote the set of indices of these rows and $\C(\A)$ to denote the linear code generated by the set of rows of $\bGN$ indexed by $\A$. Note that $\A \subseteq [N]$. The characterization of the information set $\A$ for polar codes relies on the channel polarization theorem \cite{arikan} and the concept of \textit{synthetic channel reliability}. A polar code of length $N=2^n$ is constructed by selecting a set $\A$ of indices $i\in[0,N-1]$ with high reliability \cite{arikan}. The indices in $\A$ are dedicated to information bits, while the rest of the synthetic channels with indices in $\A^c \triangleq [0,N-1]\setminus \A$ are used to transmit a known value, '0' by default, which are called \emph{frozen bits}. 
Regardless of the method we use for forming the set $\A$ for a polar code, every synthetic channel represented by index $i$, denoted by $W_n^i$ where $i\in\A$ must be more reliable than any synthetic channels in $\A^c$. 

Polar codes with the information set $\A$ can also be considered as monomial codes \cite{dragoi17thesis} where the relation between the generating set of monomials $\I\subset\M_n$ discussed in Section \ref{ssec:monmial_codes} and the information set $\A$ is as follows: 
\begin{equation}
    \forall f\in\I,\;\; \exists\, i\in\A,\; \text{ where } \supp(i)=[n]\backslash\ind(f).
\end{equation}
Accordingly, we can define set $\A_{m-r}$ equivalent to $\I_r$, collecting the row indices with $|\supp(i)|=m-r$ for $i\in\A_{m-r}$ where we have $m=n$ for polar codes while $m$ is used for decreasing monomial codes in general throughout this paper. 
The rows of matrix $\G_N={\footnotesize \begin{bmatrix}
1 & 1 \\
0 & 1
\end{bmatrix}^{\otimes n} }$ represent all possible evaluations of monomials over $\ft^n$. 
The relation between $i$ and $f$ is hence defined as 
\begin{equation}\label{eq:i_f}
    i = \sum_{j\in[n]\backslash\ind(f)} 2^j.
\end{equation}
Following \eqref{eq:i_f}, let $\barf$ defined by $\ind(\barf)=[n]\backslash\ind(f)$ represent a row of $\bG_N$. For instance, row $\bar{f}=x_4x_3x_1\rightarrow(00101)_2=5$ is equivalent to monomial $f=x_2x_0\rightarrow(11010)_2=26$. We may use $\barf$ and its decimal equivalent interchangeably. 
Note that due to the advantage of simplifying the polynomial formalism, we slightly depart from the usual convention for polar codes which is to use in the Kronecker product of $\bG_2$. Instead, we use $\G_2={\footnotesize \begin{bmatrix}
1 & 1 \\
0 & 1
\end{bmatrix} }$. It is easy to see that the two definitions (ours and the conventional one) are equivalent, they just amount to rearranging the code positions. 


\subsection{Decreasing Monomial Codes}
For the first time, it was shown in \cite{bardet2016crypt} that polar codes, similar to Reed-Muller codes, belong to a large family of codes called Decreasing Monomial Codes. Furthermore, the  monomial order was partially charatcerized in \cite{bardet} and \cite{schurch} which is called (universal) partial order in the context of sub-channels relation in terms of reliability. 
The fundamental algebraic relation between Reed-Muller codes and polar codes goes beyond their monomial structure. It is well known that both polar and Reed-Muller monomial sets ($\I$) obey a certain order relation denoted $\preceq$, as it was shown in \cite{bardet} and \cite{schurch}.

Let us define the partial order over a set of monomials. We establish the relation $\preceq$ and $\weako$ between two monomials $f$ and $g$ with the same degree and different degrees in the following. Note that the ``$w$" in $\weako$
stands for ``weak" (as in weak order) compared to $\preceq$ in the sense that any pair of monomials $f, g$ that
satisfy the relation $f \weako g$, also satisfy the relation $f \preceq g$, by definition. We will use the ``$|$" symbol to denote division between monomials, i.e., $f|g$ iff $\ind(f)\subseteq\ind(g).$ Naturally, we have the greatest common division of two monomials $\gcd(f,g)=h$ with $\ind(h)=\ind(f)\cap\ind(g).$

\begin{definition}\label{def:order}Let $m$ be a positive integer and $f,g\in\Mon.$ Then $f\weako g$ if and only if $ f|g.$ When $\deg(f)=\deg(g)=s$ say that $f\preceq_{sh} g$ if $\forall\;1\leq\ell\leq s\quad  i_\ell \le j_\ell$, where $f=x_{i_1}\dots x_{i_s}$, $g=x_{j_1}\dots x_{j_s}$. 

Define $f\preceq g\quad \text{iff}\quad \exists g^*\in \Mon\;\text{s.t.}\; f\preceq_{sh} g^*\weako g$.
\end{definition}

 \begin{remark}One might have $g^*=g$ in the definition of $\preceq.$ In such cases if $f\preceq g$ and $\deg(f)=\deg(g)$ we have $\preceq=\preceq_{sh}.$
\end{remark}

The notation $f\preceq_{sh}g$ comes the fact that one could obtain $g$ from $f$ by positively shifting some of the variables in $f.$ For example $x_2x_3\preceq_{sh}x_2x_6$ since $x_6$ is a shift by $3$ positions of $x_3.$

Remark that there is a chain relation on the variables, i.e., $x_0\preceq x_1\preceq \dots \preceq x_{m-1}.$ Also, the $\preceq$ is a order relation that is partial, e.g., $x_3x_4$ and $x_1x_5$ are not comparable with respect to $\preceq.$ The monomial sets studied in this paper are all decreasing monomial sets. 
\begin{definition}
      A set $\I \subseteq \Mon$ is \emph{decreasing}  if and only if  ($f \in \I$ and $g \preceq f $) implies
  $g \in \I$.   
\end{definition}

If we map every monomial $f\in\M_n$ to the corresponding synthetic channel denoted by $W_n^f$ in the context of polar codes, the relation between every pair of synthetic channels in terms of channel reliability can be established as follows:  Let $f$ and $g$ be two monomials such that $f\weako g$, then according to \cite[Proposition 3.3.29]{dragoi17thesis}, we have $W_n^g \preceq_{d} W_n^f$. Here, `$d$' in $\preceq_{d}$ indicates that the channel $W_n^g$ is a degradation of $W_n^f$. Now, as a polar code is generated by the set of monomials $\I\subset\M_n$, if $g\in\I$, then it implies that $f$ also belongs to $\I$.

\subsection{Permutation Group}
The set of permutations that map codewords of a code $\C$ to other codewords, i.e., leave the code invariant, forms the \emph{automorphism group} of the code $\C$, which is denoted by $\operatorname{Aut}(\C)$. Hence, A permutation $\pi$ is an automorphism of code $\C$ if and only if for every $\bc\in\C$, we have $\pi(\bc)\in\C$.

A bijective affine transformation over $\ft^m$ is represented by a pair $(\bB, \bm{\varepsilon})$ where $\bB=(b_{i,j})$ is an invertible matrix lying in the general linear group $\GL$ and $\bve$ in $\ft^m$. The action of $(\bB, \bve)$ on a monomial $g=\prod_{i\in\ind(g)}x_i$ denoted by $(\bB, \bve)\cdot g$  replaces each variable $x_i$ of $g$ by a variable $y_i$ as
\[
y_i = x_i + \sum_{j=0}^{i-1} b_{i,j} x_j + \varepsilon_i,
\]
where $b_{i,j}$ and $\varepsilon_i$ are in $\ft$. 
This new variable $y_i$, is in fact a linear form (a polynomial in which all terms have a degree at most 1). Also, the maximum variable of this linear form is $x_{i}$ as others are smaller than $x_i$ w.r.t. the order relation $\preceq.$

For decreasing monomial codes, a lower triangular affine transformation denoted by $\Alow$ is employed where $\bB\in\GL$ is a lower triangular binary matrix with $b_{i,i}=1$ and $b_{i,j}=0$ whenever $j>i$. 
Hence, the lower triangular affine group  $\Alow$ can be expressed as the following mapping from $\ft^m$ to itself. 
\[
\bx \rightarrow \bB \bx + \bve,
\]
where the matrix multiplication represents linear maps, and vector addition represents translations. For special sub-classes of decreasing monomial code, such as the Reed-Muller codes the complete permutation group is known. Indeed, the general affine group is the complete permutation group of any $\mathcal{R}(r,m)$ where $1\leq r<m-1.$  
\begin{example}
    Let $g=x_1x_4$ for $m=5$. Then we have the mapping $\bx \rightarrow \bB \bx + \bve$ as follows:
    \[
    \begin{bmatrix}
1 & 0 & 0 & 0 & 0 \\
b_{1,0} & 1 & 0 & 0 & 0 \\
0 & 0 & 1 & 0 & 0 \\
0 & 0 & 0 & 1 & 0 \\
b_{4,0} & b_{4,1} & b_{4,2} & b_{4,3} & 1 \\
\end{bmatrix} 
\begin{bmatrix}
x_0  \\
x_1 \\
x_2 \\
x_3 \\
x_4 \\
\end{bmatrix} 
+
\begin{bmatrix}
0  \\
\varepsilon_1 \\
0 \\
0 \\
\varepsilon_4 \\
\end{bmatrix} 
    \]
\end{example}
The set of polynomials resulting from the action of $\Alow$ on a monomial is collected in a set named orbit.
\begin{definition}
    The orbit of a monomial $f$ under the action of $\Alow$ is 
    defined as the set of polynomials 
    \[
        \Alow \cdot f = \{(\bB,\bve) \cdot f\mid (\bB,\bve) \in \Alow\}.
    \]
\end{definition}

Since $\Alow$ acts as a permutation on $\ev(f)$, all the elements in $\Alow\cdot f$ have the same Hamming weight. This fact will be useful when estimating the number of minimum weight codewords of a decreasing monomial code. 
\subsection{Minimum Weight Codewords}
Before focusing on larger weights, let us review how the minimum weight codewords are counted. In essence, the authors in \cite{bardet} use the action of $\Alow$ on the subset $\I_r$ where $r$ is the maximum degree of monomials in a decreasing monomial set $\I$. The problem is that when we consider the complete group $\Alow$ one needs to determine the stabilizer subgroup for each coset leader. To achieve theis goal a particular subgroup of $\Alow$ was defined.

\begin{definition}[\cite{bardet,dragoi17thesis}] For any $g\in\Mon$ define $\Alow_g$ as the subgroup
 of $(\bB,\bve) \in \Alow$ by 
\[
 \varepsilon_i = 0 \text{ if } i \not \in \ind(g) 
 ~~~\text{ and }~~ 
 b_{ij} = 
 \left \{ 
 	\begin{array}{l}
		0 \text{ if } i \not \in  \ind(g) \\
		0 \text{ if }  j \in \ind(g).
	 \end{array}
 \right.
\] 
\end{definition}

\begin{example}
    Let $g = x_0x_1$, then by definition we have $\Alow\cdot g = \{(x_0 + \varepsilon_0)(x_1 + a_{1,0}x_0 + \varepsilon_1)\mid \varepsilon_0, a_{1,0}, \varepsilon_1 \in \ft\}$. The complete group has $4$ distinct elements although as $\varepsilon_0, a_{1,0}, \varepsilon_1$ can take two values in $\{0,1\}$, we expected $2^3=8$ elements. Four repeated elements leave $g$ invariant, namely 
    $$(x_0+1)(x_1 + {\color{blue}x_0}+1) = (x_0+1)(x_1+1) = x_0x_1+x_1+x_0+1,$$ 
    $$(x_0+1)(x_1 + {\color{blue}x_0}) = (x_0+1)x_1 = x_0x_1+x_1,$$ 
    $$x_0(x_1 + {\color{blue}x_0}) = x_0(x_1 + 1) = x_0x_1 + x_0,$$
    $$x_0(x_1 + {\color{blue}x_0} + 1) = x_0x_1.$$ 
    As can be seen, since $0\in\ind(g)$, when $a_{1,0}=1$  as highlighted in blue above, the obtained polynomials are identical with another one in the group. Recall that $x_j^2=x_j$ in $\ft$. On the other hand, the subgroup $\Alow_g$ gives only $2^2$ distinct elements since by definition $a_{1,0}=0$. The elements of the subgroup are $(x_0 + 1)(x_1 + 1), (x_0 + 1)x_1, x_0(x_1 + 1),\text{ and } x_0x_1$. 
\end{example}

Notice that for any monomial $g\in \Mon$ the subgroup action of $\Alow_g$ on any monomial $f\in\Mon$ is well defined. Also, by definition of $\Alow_g$ we observe that only the variables in $g$ that are smaller, w.r.t. $\preceq$, than the variables in $f$ are making a contribution in the group action. Let us give an example.
\begin{example}
    Take $g=x_4$ and $f=x_0x_2$. Then \[\Alow_g\cdot f=\Alow\cdot f.\] 

    If we let $g=x_0x_4$ and $f=x_1x_2x_5$, then any polynomial from $\Alow_g\cdot f$ can be written as follows
    $(x_1+\varepsilon_1)(x_2+b_{2,1}x_1+\varepsilon_2)(x_5+b_{5,3}x_3+b_{5,2}x_2+b_{5,1}x_1+\varepsilon_5).$
\end{example}
What is particular about this subgroup is that it poses two major properties stated in the following theorem.

\begin{theorem}[\cite{bardet,dragoi17thesis}]\label{thm:min-weight}
Let $f\in \Mon.$ Then we have \begin{equation}
\Alow\cdot f=\Alow_f\cdot f.
\end{equation}

Moreover, there are no polynomials in $\Alow_f\cdot f$ that are fixed by more than one group element (the identity).
\end{theorem}

Now, we look at the size of the orbit $\Alow_f\cdot f$ denoted by $|\Alow_f\cdot f|$. We can break down the action of $\Alow_f$ on $f$ into two operations: 
\begin{itemize}
    \item Translation: every variable $x_i$ in the monomial can be translated by a scalar $\varepsilon_i\in\ft$ as $x_i+\varepsilon_i$. Hence, a monomial $f$ of degree $\deg(f)$, admits as many translations as possibilities for all its variables, which equals $2^{\deg(f)}$. 
    \item Linear Mapping: every variable $x_i$ can be mapped into a "new variable" ($y_i$), which is a linear combination of variable(s) $x_j$ where $j\in[i]\backslash\ind(f)$ as $y_i=x_i+\sum_{j=0, j \notin \operatorname{ind}(f)}^{i-1} b_{i,j} x_{j}$. The extra variables considered in $y_i$ express the degree of freedom we have on $x_i.$ This will be denoted by $\lambda_f(i)=|\{j\in[i] \mid j\notin\ind(f)\}|$ and represents the maximum number of variables in the group action on the variable $x_i$. The total number of free variables on all $x_i$ in the support of $f$ will be $|\lambda_f(f)|=\sum_{i\in\ind(f)}\lambda_f(i)$. 
    Then, the total possible actions for all variables in the support of $f$ is $2^{|\lambda_f|}$ (since we are defined over $\ff_2$). 
\end{itemize}
In general, we can have $\Alow_f\cdot g$, where $f$ and $g$ might be different. In such cases, we have
\begin{equation}\label{eq:lambda_general}
    |\lambda_f(g)|=\sum_{i\in\ind(g)}\lambda_f(i).
\end{equation} 
When $g=f$ we shall simplify the notations and use $\lambda_f$. 
Therefore, the cardinality of the orbit of monomial $f$  under the action of $\Alow$ is 
\begin{equation}\label{eq:A_wm}
    |\Alow_f \cdot f| = 2^{\deg(f)+|\lambda_f|}
\end{equation}

Example \ref{ex:count-min} illustrates how this counting procedure works. The previous theorem has a direct consequence, an efficient formula for counting the number of minimum weight codewords. Let us suppose we have a decreasing monomial set $\I$ with maximum degree monomials $r$, i.e., $\I_r$ is not trivial. We know that any minimum weight codeword of $\C(\I)$ is of weight $2^{m-r}$ and more significant any $\bc\in\C(\I)$ with $\w(\bc)=2^{m-r}$ has the following form $\bc=\ev(y_1\dots y_r)$ where $y_i$ are linear independent forms, i.e., $y_i\in \Rm$ with $\deg(y_i)=1$ and for any index $i$ the equation $y_i=\sum\limits_{1\leq j\leq r, j\neq i}\varepsilon_jy_j$ admits a single solution over $\ff_2$, which is the zero vector $\varepsilon_j=0,\forall j.$ The counting method relies on the following.

\begin{enumerate}
    \item Any monomial $f\in\I_r$ will define an orbit under $\Alow_f$ in which all polynomials evaluate to a minimum weight codeword; 
    \item We know how to count the cardinal of the orbit $|\Alow_f\cdot f|=|\Alow_f|$ (by Theorem \ref{thm:min-weight});
    \item For any pair of monomials $f,g\in\I_r$ the orbits $\Alow_f\cdot f$ and $\Alow_g\cdot g$ are disjoint.
    \item Finally, sum over all monomials $f\in \I_r.$
\end{enumerate}

One key ingredient used to demonstrate that the aforementioned procedure retrieves all minimum weight codewords, is the following lemma, demonstrated in Proposition 3.7.12 in \cite{dragoi17thesis}.
\begin{lemma}\label{lem:prod-lin-form-distinct-var}
    Let $P=\prod_{j=1}^{l}y_j$ be a product of $l$ independent linear forms $y_i$ each having maximum variables $x_{i_j}$. Then $P$ can be written as $P=\prod_{j=1}^{l}y_j^{*}$ where all maximum variables $x_{i_j^{*}}$ in $y_j$ are pairwise distinct.
\end{lemma}

Observe that every $f\in\I_r$ represents a coset leader for coset $\C_{\bar{f}}$ and the orbit represents the set of core rows where their row combinations along with the balancing rows result in minimum weight codewords \cite{rowshan-err_coef}.
\begin{example}\label{ex:count-min}
    Let $m=8$ and $\I$ be a decreasing monomial set with $r=4.$ Suppose $f\in\I_r=\{x_1x_0(x_3x_2,x_4x_2,x_5x_2,x_4x_3)\}$, then the table below illustrates the procedure of finding the total number of codewords with weight $\w_{min}$, i.e., $|W_{\w_{\min}}|$.
\begin{table}[!h]
        \centering
        \setlength{\tabcolsep}{2pt}
        \footnotesize
        \begin{tabularx}{\columnwidth}{c||c|c|c|c}
         \toprule
       $\ind(f)$ &$\{3,2,1,0\}$&$\{4,2,1,0\}$&$\{5,2,1,0\}$&$\{4,3,1,0\}$\\
       \midrule
    $(\lambda_f(3),\cdots,\lambda_f(0))$ &$(0,0,0,0)$& $(1,0,0,0)$ &$(2,0,0,0)$&$(1,1,0,0)$\\
    \midrule
    $|\Alow_f\cdot f|$&$2^{4}$&$2^{5}$&$2^{6}$&$2^{6}$\\
    \midrule
    $|W_{\w_{\min}}|$&\multicolumn{4}{c}{$176$}\\
    \bottomrule
        \end{tabularx}
    \end{table}
\end{example}

\section{Structural properties of codewords with Weight $1.5\wm$}


We shall begin this section by recaling a classification result on Reed-Muller codes, which will represent the foundation of our weight enumerattion. Kasami et al. characterized codewords of Reed-Muller codes with weights less than twice the minimum weight in \cite{kasami1970weight} and derived explicit formulas for the enumeration of these weights. 

\begin{theorem}{\cite{sloane1970weight},\cite[Theorem 1]{kasami1970weight}}
\label{thm:Kasami-Tokura}
 Let $r<m$ and $P\in\Rm$ be such that $\deg(P)\leq r$ with $0<\w(\ev(P)|<2^{m+1-r}.$ Then $P$ is affine equivalent (it can be transformed using an affine transformation) to one of the forms 
\begin{enumerate}
	\item $P=y_1\dots y_{r-\mu}(y_{r-\mu+1}\dots y_{r}+y_{r+1}\dots y_{r+\mu})$ where $m\geq r+\mu,r\geq \mu\geq 3$
	\item $P=y_1\dots y_{r-2}(y_{r-1}y_{r}+\dots+y_{r+2\mu-3}y_{r+2\mu-2})$ where $m-r+2\geq 2\mu, \mu\geq 2.$
\end{enumerate} 

In both cases $y_i$ are linear independent forms and $\w(\ev(P))=2^{m+1-r}-2^{m+1-r-\mu}.$
\end{theorem}

\begin{example}
    Let $m=9$ and $r=3.$ From the conditions in Thm. \ref{thm:Kasami-Tokura} we notice that $3\leq \mu\leq 3$ (case 1)) and $4\leq 2\mu\leq m-r+2=8$ which implies $\mu\leq 4$ (case 2)). Hence, have 
    \begin{itemize}
        \item $P=y_1y_2y_3$ which gives $\w(\ev(P))=\wm=2^{9-3}=64$;
        \item $\mu=2$ (case 2) in Thm. \ref{thm:Kasami-Tokura}), $P=y_1(y_2y_3+y_4y_5)$ and we have $\w(\ev(P))=2^{7}-2^{7-2}=128-32=96$;
        \item $\mu=3$ (case 1) in Thm. \ref{thm:Kasami-Tokura}), $P=y_1y_2y_3+y_4y_5y_6$ and we have $\w(\ev(P))=2^{7}-2^{7-3}=128-16=112$;
        \item $\mu=4$ (case 2) in Thm. \ref{thm:Kasami-Tokura}), $P=y_1(y_2y_3+y_4y_5+y_6y_7)$ and we have $\w(\ev(P))=2^{7}-2^{7-4}=128-8=120$; 
    \end{itemize}
\end{example}

\begin{remark}
    While for Reed-Muller codes any affine transformation globally preserves the code, in the general case of decreasing monomial codes, this fact is no longer true. Hence, for decreasing monomial codes applying directly Theorem \ref{thm:Kasami-Tokura} for counting such codewords is not possible.
\end{remark}

In this article we will restrict our analysis to the case $\mu=2$ which is equivalent to codewords of weight $1.5\wm.$ 

\begin{corollary}\label{cor:Kasami-cor1}
Let $r<m$ be positive integers s.t. $r=\max_{\ev(P)\in\C}\deg(P)$. Then any codeword of weight $1.5\wm$, up to an affine transformation, is equal to $\ev(y_1\dots y_{r-2}(y_{r-1}y_{r}+y_{r+1}y_{r+2})).$
\end{corollary}

Tacking a closer look at the shape of the polynomial in Corollary \ref{cor:Kasami-cor1} we deduce that any codeword of weight $1.5\wm$ is equal to the sum of two minimum weight codewords. Indeed, both $y_1\dots \dots y_{r}$ and $y_1\dots y_{r-2}y_{r+1}y_{r+2}$ define minimum weight codewords (as product of $r$  independent linear forms), and thus they belong to two distinct orbits $\Alow\cdot f, \Alow\cdot g.$ Hence, our first task is to understand what is the structure of the sum of these two sets. 

Recall that the subgroup $\Alow_f$ on $f$ generates the exact same orbit as the complete group action $\Alow$ on any monomial $f.$ Since here we are dealing with sums and product of orbits we would like to know how the subgroup property is preserved in this case.
\begin{lemma}\label{lem:decomp-ltam-gcd}Let $f,g\in\I_r$ and $h=\gcd(f,g)\in\Mon.$ Then 
\begin{multline*}
    \Alow\cdot h\cdot \left(\Alow\cdot \frac{f}{h}+\Alow\cdot \frac{g}{h}\right)\\=
    \Alow_h\cdot h\cdot \left(\Alow_{f}\cdot \frac{f}{h}+\Alow_g\cdot \frac{g}{h}\right).
\end{multline*}
    
\end{lemma}
 The proof of our result can be found in Appendix \ref{app:proof-lem:decomp-ltam-gcd}. Let us give an example on how this decomposition works.
\begin{example}
    Let $h=x_0x_6, f/h=x_3x_2$ and $g/h=x_5x_1.$ The polynomial $P=x_0(x_6+x_2)\left((x_3+x_0)x_2+x_5(x_1+x_0)\right)$ belongs to $ \Alow\cdot h\cdot \left(\Alow\cdot \frac{f}{h}+\Alow\cdot \frac{g}{h}\right)$. By expanding the product we have 
    \begin{align*}
        P&={\color{blue}x_0}(x_6+x_2)(x_3+{\color{blue}x_0})x_2+{\color{blue}x_0}(x_6+x_2)x_5(x_1+{\color{blue}x_0})\\
        &={\color{blue}x_0}(x_6+x_2)(x_3+{\color{blue}1})x_2+{\color{blue}x_0}(x_6+x_2)x_5(x_1+{\color{blue}1})\\
        &=x_0(x_6+x_2)\left((x_3+1)x_2+x_5(x_1+1)\right)
    \end{align*}
    So, $P$ is an elements of the set $ \Alow_h\cdot h\cdot \left(\Alow_{f}\cdot \frac{f}{h}+\Alow_g\cdot \frac{g}{h}\right).$
\end{example}

 Moving forward, we propose a classification theorem for codewords of weight $1.5\wm$ for decreasing monomial codes. This result is somehow the equivalent of Kasami and Tokuras's classification theorem where affine transformations are allowed, to a classification where transformations from the $\Alow$ are allowed.

\begin{theorem}\label{thm:equality-orbits}
    Let $\C(\I)$ be a decreasing monomial code and $r=\max_{f\in \I}\deg(f)$. Then any codeword of weight $1.5\wm$, say $\ev(P)$ is s.t. $\exists f,g\in \I_r$ with $P\in\Alow_h\cdot h\cdot\big(\Alow_{f} \cdot \frac{f}{h}+\Alow_{g}\cdot \frac{g}{h}\big)$, where  $h=\gcd(f,g)$, and $\deg(h)=r-2$.
\end{theorem}

 Notice that this result is constructive, i.e., given a codeword $\bc=\ev(P)$ of weight $1.5\wm$, one can compute two monomials $f,g$ such that $P$ belongs to the Minkowski sum of their orbits under $\Alow.$ In the following example we will illustrate how to compute the monomials $f,g$ given $P$ (for more details see the proof of the theorem in Appendix \ref{app:proof-thm:equality-orbits}).

         
\begin{example}
    Let $m=9,r=3.$ Any $P\in\Rm$ with $\w(\ev(P))=2^{9+1-3}-2^{9+1-5}=96=1.5\times 64=1.5\wm.$ Consider the following cases. 
    \begin{enumerate}
        \item one factor, all distinct maximum variables:\\ $P=({\color{blue}x_4}+x_0){\color{blue}x_2x_1}+({\color{blue}x_5}+x_1)({\color{blue}x_4}+x_0){\color{blue}x_3}$. By Thm. \ref{thm:Kasami-Tokura} $\ev(P)$ is a $1.5\wm$ since once $P$ is factored we obtain $P=({\color{blue}x_4}+x_0)({\color{blue}x_2x_1}+({\color{blue}x_5}+x_1){\color{blue}x_3})$, which has the form $y_1(y_2y_3+y_4y_5).$ By simply tacking the maximum variables in each $y_i$ we can set $f=x_4x_2x_1,g=x_5x_4x_3$ and hence $h=\gcd(f,g)=x_4,f/h=x_2x_1,g/h=x_5x_3.$ 
         \item one factor, four distinct maximum variables:\\ $P=({\color{blue}x_4}+x_0)({\color{blue}x_5}+x_2){\color{blue}x_1}+({\color{blue}x_5}+x_1)({\color{blue}x_4}+x_0){\color{blue}x_3}$, once factored, as in the previous case, it leads to the required form in Theorem \ref{thm:Kasami-Tokura}, $P=y_1(y_2y_3+y_4y_5).$ However, simply setting $f,g$ using the maximum variables does not work. Indeed, if would lead to $f=x_5x_4x_1,g=x_5x_4x_1$, which is not the form stated in Thm. \ref{thm:equality-orbits}. However, reshaping the terms in $P$ as in the proof of Thm. \ref{thm:equality-orbits}, one gets 
         $P=({\color{blue}x_4}+x_0)(({\color{blue}x_5}+x_1)({\color{blue}x_3}+x_1)+({\color{blue}x_2}+x_1){\color{blue}x_1}).$ Now, taking the maximum variables in each linear independent form leads to $f=x_5x_4x_3,g=x_4x_2x_1$, as required in Thm. \ref{thm:equality-orbits}. 
          \item one factor, three distinct maximum variables:\\ $P=({\color{blue}x_6}+x_0)({\color{blue}x_5}+x_4)({\color{blue}x_3}+x_1+1)+({\color{blue}x_5}+x_0)({\color{blue}x_6}+x_0)({\color{blue}x_3}+1)$. Indeed, by Theorem \ref{thm:Kasami-Tokura}, this corresponds to a $1.5\wm$-weight codeword, since one can write $P=y_1(y_2y_3+y_4y_5)$ where $y_i$ are all linear independent forms. If we simply select the maximum variables (in blue) in each $y_1$ we would get $f=g=x_6x_5x_3$ which is not really helpful. Cleverly manipulating the polynomial, one obtains 
         $P=({\color{blue}x_6}+x_0)(({\color{blue}x_5}+x_0){\color{blue}x_1}+({\color{blue}x_4}+x_0)({\color{blue}x_3}+x_1+1))$, from which $f=x_6x_5x_1, g=x_6x_4x_3$, exactly as in Thm. \ref{thm:equality-orbits}.
          
    \end{enumerate}
\end{example}
For Reed-Muller codes Theorem \ref{thm:Kasami-Tokura} combined with the general affine group were the two main ingredients for characterizing and counting codewords of weight smaller than $2\wm.$ Decreasing monomial codes do not admit the complete general affine group. Hence, our result provides the first step towards understanding how such codewords are formed when we deal with decreasing monomial codes.

Going further, we will determine the cardinality of $\Alow_h\cdot h\cdot\big(\Alow_f\cdot \frac{f}{h}+\Alow_{g}\cdot \frac{g}{h}\big)$ from Theorem \ref{thm:equality-orbits}. The major challenge here is to estimate $|\Alow_{f} \cdot \frac{f}{h}+\Alow_{g}\cdot \frac{g}{h}|.$ It is obvious that the following upper bound \cite{tao2006book} holds $|\Alow\cdot f+\Alow\cdot g|\leq  |\Alow\cdot f||\Alow\cdot g|.$
However, reaching the upper bound is not always the case. Let us start by giving an example.

\begin{example}\label{ex:cardinality-orbits}
Let $f=x_4x_2x_0$ and $g=x_3x_2x_1.$ Then, any polynomial $P\in\Alow\cdot f$ and $Q\in\Alow\cdot g$ can be written as $P=(\bA,\bve)\cdot f,Q=(\bB,\bvg)\cdot g$, or equivalently
\begin{align*}
    P&=(x_4+b_{4,3}x_3+b_{4,1}x_1+\varepsilon_4)(x_2+b_{2,1}x_1+\varepsilon_2)(x_0+\varepsilon_0)\\
    Q&=(x_3+a_{3,0}x_0+\gamma_3)(x_2+a_{2,0}x_0+\gamma_2)(x_1+a_{1,0}x_0+\gamma_0)
\end{align*} where the scalars $a_{i,j},b_{i,j},\varepsilon_{i},\gamma_{i}\in\ff_2.$ 

The cardinality of the orbits are $|\Alow\cdot f|=2^{3+(2+1+0)}=2^6$ and $|\Alow\cdot g|=2^{3+(1+1+1)}=2^6$. Hence, we have \begin{equation*}
    |\Alow\cdot f+\Alow\cdot g|\leq 2^{12}.
\end{equation*}
\end{example}
In this example, however, the upper bound is not achieved due to the existence of \emph{collisions}. Let us define the term collision as follows:
\begin{definition}\label{def:collision}
    Collision: Any pair of distinct polynomials $P,P^{*}\in\Alow\cdot f$ and $Q,Q^{*}\in\Alow\cdot g$ with $P\neq P^{*}, Q\neq Q^{*}$ such that they produce the same sum, that is, $P+Q=P^{*}+Q^{*}$, results in a collision. 
\end{definition}

An example for collision is $P\!=\!(x_4+x_3)x_2x_0, Q\!=\!x_3x_2x_1$ and $P^{*}\!=\!x_4x_2x_0, Q^{*}\!=\!x_3x_2(x_1+x_0).$ 
Notice that in Example \ref{ex:cardinality-orbits}, we were able to generate an invariant for $x_4x_2x_0+x_3x_2x_1$ by simply inserting in each monomial a variable from the other monomial. It is not always possible to do so. 

We need to define the number of collision for two monomials. 
\begin{definition} Let $f=x_{i_1}x_{i_2}$ and $g=x_{j_1}x_{j_2}$ with $\gcd(f,g)=1$ and $i_2>j_2.$ The \emph{degree of collision} of $f$ and $g$ is\\
    \[\alpha_{{f},{g}}=\left\{
  \begin{array}{cc}
       0& i_2>i_1>j_2>j_1  \\
       1& i_2>j_2>i_1>j_1\\
       2& i_2>j_2>j_1>i_1
  \end{array}
  \right..\]
\end{definition}

For any given pair $(P,Q)$ with $P\in\Alow\cdot f$ and $q\in\Alow\cdot g$, the quantity ${\alpha_{f,g}}$ will allow us to count how many collisions we get for any fixed pair of polynomials. Indeed, as we will demonstrate, this parameter in nothing more that the number of independent linear equations one needs to satisfy for collisions given a pair of monomials $f,g.$ Hence, it is normal that there is a connection between the structure of $f$ and $g$ and this parameter. 
\begin{example}
    Let's consider two cases separately
    \begin{itemize}
    \item $i_2>j_2>j_1>i_1$ with $f=x_6x_2,g=x_5x_3$, which imply $\alpha=2.$ For $P=(x_6+x_4+1)(x_2+x_1)\in \Alow\cdot f$ and $Q=(x_5+x_4+x_2)(x_3+x_2+x_0+1)\in\Alow\cdot g$. We can create two non-trivial distinct pairs $P^{*},Q^{*}$ 
    \begin{itemize}
        \item $P^{*}=(x_6+x_4+x_3+x_2+x_0)(x_2+x_1)$ and $Q^{*}=(x_5+x_4+x_1)(x_3+x_2+x_0+1)$ 
        \item $P^{*}=(x_6+{\color{blue}x_5}+x_3+x_2+x_1+x_0)(x_2+x_1)$ and $Q^{*}=(x_5+x_4+x_1)(x_3+x_1+x_0+1)$
    \end{itemize}
    \item  $i_2>j_2>i_1>j_1$ with $f=x_4x_2,g=x_3x_0.$ We obtain that $f+g=(x_4+x_0)x_2+(x_3+x_2)x_0.$
    \end{itemize}
\end{example}

Why is the parameter $\alpha_{f,g}$ important? Mainly because it will help us count how many polynomials are overcounted.  
\begin{proposition}\label{pr:degree-two-minkowski}
     Let $\I\subseteq\Mon$ be a decreasing monomial set and $f=x_{i_1}x_{i_2}$ and $g=x_{j_1}x_{j_2}$ with $\gcd(f,g)=1$ and $i_2>j_2.$ Then 
\begin{multline}
    |\Alow\cdot f+\Alow\cdot g|=\\ \frac{|\Alow\cdot f|\times |\Alow\cdot g|}{2^{\alpha_{{f},{g}}}},
\end{multline}
\end{proposition}
The proof of this result is provided in the appendix \ref{app:proof-pr:degree-two-minkowski}.

While up to this point we have characterized the cardinality of a Minkowski sum of distinct orbits, there is still a last question left unanswered. What is the cardinality of the product $\Alow_h\cdot h\cdot\big(\Alow_{f} \cdot \frac{f}{h}+\Alow_{g}\cdot \frac{g}{h}\big).$ The answer here is rather obvious, the cardinal of the product set is equal to the product of the cardinal of each set. 
\begin{lemma}\label{lem:cardinal-product-orbits}
    Let $\I$ be a decreasing set with $r=\max_{f\in\I}\deg(f)$ and $f,g\in \I_r$ with $h=\gcd(f,g)$ s.t. $\deg(h)=r-2.$ Then \begin{multline}
        |\Alow_h\cdot h\cdot\big(\Alow_{f} \cdot \frac{f}{h}+\Alow_{g}\cdot \frac{g}{h}\big)|=\\
        |\Alow_h\cdot h|\times |\big(\Alow_{f} \cdot \frac{f}{h}+\Alow_{g}\cdot \frac{g}{h}\big)|
    \end{multline}
\end{lemma}
\section{Counting formula for  $1.5\wm$ weight codewords }\label{sec:counting-formula}


We have previously demonstrated that any decreasing monomial code with monomial set $\I$ and maximum degree monomials $r=\max_{f\in \I}\deg(f)$ will contain codewords of weight $1.5\wm$ if and only if there are monomials of degree $r$ (denote a pair of such monomials $(f,g)$) such that they share $r-2$ variables, i.e., $\deg(\gcd(f,g))=r-2.$ For example, a decreasing monomial code with maximum degree monomials $x_0x_1x_2,x_0x_1x_3$ will not have codewords of weight $1.5\wm$, while the code with maximum degree variables $x_0x_1x_2,x_0x_1x_3,x_0x_2x_3,x_0x_1x_4$ will have codewords of weight $1.5\wm$ since the pair $(x_0x_1x_4,x_0x_2x_3)$ has the common factor $x_0$ of degree $r-2=1$.

Hence, from Theorem \ref{thm:equality-orbits} we deduce the following.
\begin{corollary}\label{cor:1.5d}
    Let $\I$ be a decreasing monomial set with $r=\max_{f\in\I}(\deg(f))$. Then we have \begin{multline}
        W_{1.5\wm}=\bigcup\limits_{\substack{f,g\in\I_r\\ h=\gcd(f,g)\in \I_{r-2}}}\Alow_h\cdot h\\\cdot\left(\Alow_{f}\cdot \frac{f}{h}+\Alow_{g} \cdot \frac{g}{h}\right)
    \end{multline}      
\end{corollary}

Next, we will demonstrate that for distinct monomial pairs $(f,g),(f^{*},g^{*})$ define disjoint polynomial sets. In a sense, we transpose the property of disjoint orbits from minimum weight codewords to $1.5\wm.$ 
\begin{proposition}\label{pr:no-colision-1.5dorbits}
    Let $\I$ be a decreasing monomial set and let $(f,g),(f^{*},g^{*})\in \I_r\times\I_r$ with $(f,g)\neq(f^{*},g^{*})$ and $\deg(h)=\deg(h^{*})=r-2$, where $h=\gcd(f,g)$ and $h^{*}=\gcd(f^{*},g^{*}).$ Then the sets $\Alow_h\cdot h\cdot(\Alow_{f} \cdot \frac{f}{h}+\Alow_{g}\cdot \frac{g}{h})$ and $\Alow_{h^{*}}\cdot h^{*}\cdot(\Alow_{f^{*}} \cdot \frac{f^{*}}{h^{*}}+\Alow_{g^{*}}\cdot \frac{g^{*}}{h^{*}})$ are disjoint.
\end{proposition}


We are now in possession of three ingredients that mixed together will give us a closed formula for $1.5\wm$ weight codewords of any decreasing monomial code, and thus, implicitly of polar and ReedMuller codes. Let us recall them and then state our result.
\begin{itemize}
    \item any $1.5\wm$ weight codeword is the evaluation of a polynomial that belongs to a Minkowski sum of two orbits (Theorem \ref{thm:equality-orbits})
    \item two distinct pair of degree $r$ monomials generate distinct orbits (Proposition \ref{pr:no-colision-1.5dorbits})
    \item we know how to count the cardinality of Minkowski sums of orbits (Proposition \ref{pr:degree-two-minkowski})
\end{itemize}  
\begin{theorem}\label{thm:formula_15w}Let $\I$ be a decreasing monomial set and $r=\max_{f\in \I}\deg(f).$
  \begin{equation}\label{eq:formula_15w}
      |W_{1.5\wm}|=\sum\limits_{\substack{f,g\in \I_r\\ h=\gcd(f,g)\in\I_{r-2}}} 2^{r+2+|\lambda_h|+|\lambda_{f}(\frac{f}{h})|+|\lambda_{g}(\frac{g}{h})|-\alpha_{\frac{f}{h},\frac{g}{h}}}
  \end{equation} 
  where $\frac{f}{h}=x_{i_1}x_{i_2},\frac{g}{h}=x_{j_1}x_{j_2}$ with $i_2>j_2.$
\end{theorem}


The closed form expression proposed in Theorem \ref{thm:formula_15w} for counting the codewords with weight $1.5\wm$ can be implemented by a simple procedure. A MATLAB script realising equation \eqref{eq:formula_15w} can be found in Appendix \ref{app:matlab}. Observe that the procedure has the complexity of order $O(|\I_r|^2\cdot r)$ due to the nested loops. A MATLAB script is available in \cite{github}. 
\begin{example}\label{ex:enum-1.5d}
    Let us consider the polar code (128,64) as decreasing monomial code with $m=7$ and $r=4.$ Let the set $\I_r$ of monomials of degree $4$ be defined by
$$x_0x_1x_2x_3,\;x_0x_1x_2x_4,\;x_0x_1x_3x_4,$$
$$x_0x_2x_3x_4,\;x_1x_2x_3x_4,\;x_0x_1x_2x_5,\;x_0x_1x_3x_5$$
equivalent to $\A_{m-r}=\{112,104,100,98,97,88,84\}$ as row indices of $\bG_N$. 
Table~\ref{tab:ex-1.5d} shows all the possible combinations of monomials that admit a degree $r-2=2$ common factor $h$ and their associated cardinalities. 
The penalties account for the collisions, i.e., the identical codewords that should not be counted. The column "Total" results from the multiplication of the cardinalities of all subgroups and the penalties. 

\begin{table*}[]
    \centering
    \begin{tabular}{c|c||c|c|c||c|c|c|c||c}
         \toprule
         $\bar{f},\bar{g}$& $\ind(f),\ind(g)$&$\ind(h)$&$\ind(\frac{f}{h})$&$\ind(\frac{g}{h})$&$|\Alow_h\!\cdot\! h|$&$|\Alow_f\!\cdot\! \frac{f}{h}|$&$|\Alow_g\!\cdot\! \frac{g}{h}|$& Penalties &Total \\
         \midrule
    104, 84 & $[ 0, 1, 2, 4 ], [ 0, 1, 3, 5 ]$& $[0,1]$& $[2,4]$&$[3,5]$&$2^{2+0+0}$&$2^{2+0+1}$&$2^{2+1+2}$&$2^{-1}$&512  \\
         \midrule
    \color{blue}100, 88 & $[ 0, 1, 3, 4 ], [ 0, 1, 2, 5 ]$&$[0,1]$&$[3,4]$&$[2,5]$&$2^{2+0+0}$&$2^{2+1+1}$&$2^{2+0+2}$&$2^{-2}$& \color{blue}256\\
    \midrule
   \color{blue}98, 88 & $[ 0, 2, 3, 4 ], [ 0, 1, 2, 5 ]$&$[0,2]$&$[3,4]$&$[1,5]$&$2^{2+0+1}$&$2^{2+1+1}$&$2^{2+0+2}$&$2^{-2}$& \color{blue}512\\
    \midrule
   98, 84 & $[ 0, 2, 3, 4 ], [ 0, 1, 3, 5 ]$&$[0,3]$&$[2,4]$&$[1,5]$&$2^{2+0+2}$&$2^{2+1+1}$&$2^{2+0+2}$&$2^{-2}$& 1024\\
     \midrule
   97, 88 & $[ 1, 2, 3, 4 ], [ 0, 1, 2, 5 ]$&$[1,2]$&$[3,4]$&$[0,5]$&$2^{2+1+1}$&$2^{2+1+1}$&$2^{2+0+2}$&$2^{-2}$& 1024\\
     \midrule
   97, 84 & $[ 1, 2, 3, 4 ], [ 0, 1, 3, 5 ]$&$[1,3]$&$[2,4]$&$[0,5]$&$2^{2+1+2}$&$2^{2+1+1}$&$2^{2+0+2}$&$2^{-2}$& 2048\\
     \bottomrule
    \end{tabular}
    \caption{Illustration of enumerating the polar code (128,64) given in Example \ref{ex:enum-1.5d}, where $A_{1.5\wm}=|W_{1.5\wm}|=5376$.} 
    \label{tab:ex-1.5d}
\end{table*}
\end{example}

\begin{example}\label{ex:enum-d-1.5d}
Let us take the polar code (128,64) in Example \ref{ex:enum-1.5d}. Since the construction of polar codes is channel dependent, one can improve the code by reducing the size of set $\I_r$ (or equivalently by adjusting the design-SNR using any construction method \cite{vangala}). Table \ref{tab:ex-Ir-change} shows the sets $\A_{m-r}$ corresponding to design-SNR = 0, 3, 6, and 8, from top to bottom. 
Note that when $\A_{m-r}=\{112,104\}$, or equivalently $\I_r=\{x_0x_1x_2x_3,x_0x_1x_2x_4\}$, the common factor $x_0x_1x_2$ is of degree $3 > r-2$ that does not satisfy the condition to have $1.5\wm$-weight codewords. Hence, no codewords with 1.5$\wm$ exist. Moreover, by changing the design-SNR, we are moving from polar codes to Reed-Muller codes. 

As an example for longer codes, let us take the polar code (2048,1024) where $m=11$. The last two rows of Table \ref{tab:ex-Ir-change} show the results for the enumeration of this code constructed with design-SNR = 2 and 3 (bottom). 
\begin{table}[!ht]
    \centering
    \setlength{\tabcolsep}{3pt}
    \renewcommand{\arraystretch}{1.5}
    \begin{tabular}{l|l|l|l|l|l|l}
        \toprule
        $\A_{m-r}$ & $m$ & $r$ & $\wm$ & $A_{\wm}$ & $A_{1.5\wm}$ & Ref. \\ \hline
        $\{112,104,{\color{blue}100,98},97,{\color{blue}88},84\}$ & 7 & 4 & 8 & 688 & 5376 & \\ \hline
        $\{112,104,{\color{blue}100,98,88}\}$ & 7 & 4 & 8 & 304 & \color{blue}768 & \\ \hline
        $\{112,104\}$ & 7 & 4 & 8 & 48 & 0 & \cite{yao}\\ \hline
        $\{120,116,\cdots, 23, 15\}$ & 7 & 3 & 16 & 94488 & 74078592 & \cite{sugino}\\ \midrule
        $\{1920,1856,\cdots,1680\}$ & 11 & 7 & 16 & 11648 & 215040 & \cite{li_adaptive}\\ \hline
        $\{1920,1856\}$ & 11 & 7 & 16 & 384 & 0 & \\ 
        \bottomrule
    \end{tabular}
    \caption{The Impact of code constructions, i.e., set $\A$, on $A_{1.5\wm}$ for codes (128,64) and (2048,1024) discussed in Example \ref{ex:enum-d-1.5d}. Our results match the results in the provided references.}
    \label{tab:ex-Ir-change}
    \vspace{-10pt}
\end{table}
Since the enumeration of codewords with weight $\wm$ and $1.5\wm$ depends on set $\I_r$, a reduction in the cardinality of this set can reduce both. This can be observed in Table \ref{tab:ex-Ir-change}.
\end{example}

\section{On the Formation of 1.5$\wm$-weight Codewords via $\bG_N$-Row Combinations}
In this section, we show the implications of discussions in the previous sections on the formation of 1.5$\wm$-weight codewords via combining particular pairs of $\wm$-weight codewords. 
As shown in \cite{rowshan22ErrCoef,rowshan22EnuPAC}, $\wm$-weight codewords of a polar code $\C(\A)$ are formed in the cosets led by a row $\bgi, i\in \Amr$. However, the 1.5$\wm$-weight codewords are generated by combining $\wm$-weight codewords of distinct cosets. 
Hence, let us first define the cosets as following:
\begin{definition}\label{def:cosets}
Cosets: we can partition a polar code into $|\A|$ disjoint cosets $\Ci(\A)=\bgi+\C(\A \setminus [0,i])$ of its subcodes $\C(\A \setminus [0,i])$ for $i\in \A$ where $\bgi$ is the $i$-th row of the polar transform $\bGN$, that is
\begin{equation}
    \Ci(\A) \triangleq \left\{\bgi\oplus\bigoplus_{h\in \H_i} \bgh \colon \H_i \subseteq \A \setminus [0,i]\right\}\subseteq \C(\A).
\end{equation}
\end{definition}
Note that a coset led by row $\barf$ can be considered as the  action of permutation group $\Alow_f$ on $f$. 
As the weight of the codewords in every coset $\CiI$ is \cite{rowshan-err_coef}
\begin{equation}\label{eq:geq_wi}
    \w(\bgi\oplus\bigoplus_{j\in\H_i}\mathbf{g}_h)\geq \w(\bgi),
\end{equation}
where $\H_i\subseteq [i+1,N-1]$, the weight of codewords of the code $\C(\A)$ lying in the coset $\Ci(\A)$ for any $i\not\in\A_{m-r}$ will be larger than $\wm$. Hence, we only consider the cosets $\Ci,i\in\Amr$ for $\wm$-weight codewords. 

The indices of core rows in  $\bG_N$ forming $\wm$-weight codewords in coset $\C_{\barf}$ are collected in set $\K_{\barf}$ defined as
\[
    \K_{\barf} = \{i\in\A_{m-r}\backslash[0,\barf]\colon |\ind(i)\backslash\ind(\barf)|=1\},
\]
and $|\K_{\barf}|=\deg(f)+|\lambda_f|$. Recall the relation between $\barf$ and $f$ as the binary representation of $\bar{f}$ is 1's complement of the binary representation of $f$. 
Suppose we have two $\wm$-weight codewords $\bc_{\barf}$ and $\bc_{\barg}$, equivalent to $\ev(P)$ and $\ev(Q)$ where  $P\in\Alow_{f}\cdot f$ and $Q\in\Alow_{g}\cdot g$. The codeword $\bc_{\barf}$ is generated by  
\begin{equation}\label{eq:decomposition}
\bc_{\barf} = \bg_{\barf} \oplus \bigoplus_{j\in\J}\bgj \oplus \bigoplus_{m\in\M(\J)}\bgm.
\end{equation}
where $\J\subseteq \K_{\barf}$ and $\M(\J)$ is defined as a set of additional rows to get $\w(\bc_{\barf})=\wm$ if $\w\big( \bg_{\barf} \oplus \bigoplus_{j\in\J}\bgj \big)>\wm$ (see \cite{rowshan22EnuPAC}). The codeword $\bc_{\barg}$ is similarly formed. 

To get a 1.5$\wm$-weight codeword by adding $\bc_{\barf}$ and $\bc_{\barg}$, i.e., $\w (\bc_{\barf} \oplus \bc_{\barg}) = 1.5 \wm$, codewords $\bc_{\barf}, \bc_{\barg}$ must satisfy the following condition:
\[
    |\ind(\bc_{\barf})\cap\ind(\bc_{\barg})| = \frac{1}{4}\wm.
\]
These codewords exist in cosets $\C_{\barf}$ and $\C_{\barg}$ where 
\begin{equation}\label{eq:cond_1.5w}
    |\ind(\barg)\backslash\ind(\barf)|=2.
\end{equation}
This is equivalent to Corollary \ref{cor:Kasami-cor1} where there exist two distinct variables in each term, i.e., $y_{r-1}y_{r}$ and $y_{r+1}y_{r+2}$. To keep this condition satisfied for every pair of codewords in distinct cosets, we need to limit the choice of codewords from each coset to the ones matching in $|\ind(f)\cap\ind(g)|=r-2$ shared variables. This reduces the choice of codewords from $2^{|\K_{\barf}|}$ to $2^{|\K_{\barf}|-(r-2)}$ in $\C_{\barf}$ or alternatively in $\C_{\barg}$. That is, we do not have the freedom to combine any pair of $\wm$-codewords from the cosets but they should match.  In terms of row combinations in each coset, this means the values of coordinates $\ind(f)\cap\ind(g)$ in the binary representation of row indices in each coset should match the rows in the other cosets. 
\begin{example}
    Let us take cosets $\C_{84}$ and $\C_{104}$ in the polar code (128,64) where $m=7,r=4$. Observe that rows $\bg_{104}$ and $\bg_{84}$ are individually considered $\wm$-weight codewords and $\ind(f)\cap\ind(g)=\{0,1\}$. Since the condition \eqref{eq:cond_1.5w} is satisfied for $\barf=104,\barg=84$, we have $\w(\bg_{104}\oplus\bg_{84})=1.5\wm$. Now, we add row $85=(10101{\color{blue}01})_2\in\K_{84}$ to $\bg_{84}$ to form another $\wm$-weight codeword in coset $\C_{84}$. however, to form $1.5\wm$-weight codeword in combination with $\bg_{104}$, we have to include row $105=(11010{\color{blue}01})_2\in\K_{104}$ in coset $\C_{104}$ as well. Observe that the binary digits at coordinates 0,1 (highlighted in blue) for the rows of two cosets are matched and as a result  $\w((\bg_{104}\oplus\bg_{105})\oplus(\bg_{84}\oplus\bg_{85}))=1.5\wm$.
\end{example}
Hence, the number of $1.5\wm$-weight codewords resulting from the combination of every pair of $\wm$-weight codewords, one from $\C_{\barf}$ and the other from $\C_{\barg}$, can be at most
$A_{1.5\wm}\{\barf,\barg\}\leq 2^{|\K_{\barf}|}\times 2^{|\K_{\barg}|}\times 2^{-(r-2)}$. 
Moreover, when $\K_{\barf}\cap \K_{\barg}\not=\emptyset$, we will have multiple identical codewords, equivalent to collisions in Definition \ref{def:collision}. 
\begin{example}
    Let us take $\barf,\barg$ as $22,25\in\A_{m-r}$ for the polar code (32,16) where $m=5,r=2$. Since $\K_{\barf}\cap \K_{\barg}=\{26,28\}$, the rows $\bg_{26}$ and $\bg_{28}$ involved in the formation of $\wm$-weight codewords in both cosets $\C_{22}$ and $\C_{25}$, e.g., $\w(\bg_{22}\oplus\bg_{26})=\wm$ in $\C_{22}$ and $\w(\bg_{25}\oplus\bg_{26})=\wm$  in $\C_{25}$. As a result, the overcounting occurs if we consider the codewords $(\bg_{22}\oplus{\color{blue}\bg_{26}})\oplus\bg_{25}$ and $\bg_{22}\oplus(\bg_{25}\oplus{\color{blue}\bg_{26}})$ distinct while they are not.
\end{example}
Hence, there will exist $2^{|\K_{\barf}\cap \K_{\barg}|}$ redundant codewords and we need to avoid collisions by discounting them as
\begin{equation}\label{eq:A_1.5w}
    A_{1.5\wm}\{\barf,\barg\} = 2^{|\K_{\barf}|+|\K_{\barg}|-(r-2)-|\K_{\barf}\cap \K_{\barg}|}.
\end{equation}
Note that \eqref{eq:A_1.5w} is equivalent to the summation terms in \eqref{eq:formula_15w}.

\begin{example}
    Let us take $\barf,\barg$ as $22,25\in\A_{m-r}$ for the polar code (32,16) where $m=5,r=2$. Observe that $|\ind(\barg)\backslash\ind(\barf)|=|\{0,3,4\}\backslash\{1,2,4\}|=2$, $\K_{22}=\{23, {\color{blue}26}, {\color{blue}28}, 30\}$, and $\K_{25}=\{{\color{blue}26}, 27, {\color{blue}28}, 29\}$. As  $\ind(g)\cap\ind(f)=\{0,3\}\cap\{1,2\}=\emptyset$, then we have $A_{1.5\wm}\{\barf,\barg\} = 2^{4}\times 2^{4}\times 2^{-2}=64$. Observe that since we have $r=2$ and $|\ind(\barg)\backslash\ind(\barf)|=2$ for every pair in $\A_{m-r}$, the condition $\ind(h)=\ind(g)\cap\ind(f)=\emptyset$ is always satisfied. As another example, let us take $\barf,\barg$ as $104,84\in\A_{m-r}$ for the polar code (128,64) where $m=7,r=4$ (see 1st row of Table \ref{tab:ex-1.5d}). Since $|\K_{104}|=5$, $|\K_{84}|=7$, and $\K_{104}\cap\K_{84}=\{112\}$, then $A_{1.5\wm}\{104,84\} = 2^{5}\times 2^{7}\times 2^{-2} \times 2^{-1}=512$.
\end{example}

\section{Conclusion}
This paper presents a framework for the characterization of codewords of decreasing monomial codes with weights less than $2\wm$. Specifically, we provide a closed-form expression for the enumeration of $1.5\wm$-weight codewords. We also demonstrate how $1.5\wm$-weight codewords are formed by combining a pair of $\wm$-codewords. The results show that the number of $1.5\wm$-weight codewords depends solely on the maximum-degree monomials.





\printbibliography

\appendices
\section{MATLAB Script for Enumeration of Codewords with Weight $\wm$ and $1.5\wm$} \label{app:matlab}

The following listing presents the MATLAB\textsuperscript{\texttrademark} function \emph{weight\textunderscore enum}, which serves to enumerate the first two minimum weight codewords. The inputs required are the set $\A$ of coordinates of the non-frozen bits and $n=\log_2(N)$ where $N$ is the code length.  It is worth noting that the elements of set $\A$ must be sorted in ascending order. This is crucial, as the conditions for determining $\alpha_{\frac{f}{h},\frac{g}{h}}$ are dependent on the order of $f$ and $g$, as detailed in Theorem \ref{eq:formula_15w}. 
The function begins by extracting the maximum-degree monomials from set $\A$ and storing them in set $\Amr$ (lines 2 and 3). 
Next, in the two inner for-loops, the function finds all distinct pairs of monomials in set $\Amr$ named $f$ and $g$, whose greatest common divisor (GCD) has a degree of $r-2$. Utilizing \eqref{eq:A_wm}, \eqref{eq:formula_15w}, and the function \emph{lambda} representing \eqref{eq:lambda_general}, $A_{\w}$ for $\w\in\{\wm,1.5\wm\}$ is computed and returned. 
{\footnotesize
\begin{lstlisting}
function [w,A_w] = weight_enum(A,n)
    r = max(sum(~(dec2bin(A)-'0'),2));
    Amr = A(find(sum(~(dec2bin(sort(A))-'0'),2)==r))
    Amr_sub = Amr; w=zeros(1,2); A_w=zeros(1,2);  w(1)=2^(n-r); w(2)=1.5*w(1);
    for i = Amr
        f = find(~(reverse(dec2bin(i,n))-'0'));
        A_w(1) = A_w(1) + 2^(r+lambda(f,f));
        Amr_sub(Amr_sub==i) = [];
        for j = Amr_sub
            g = find(~(reverse(dec2bin(j,n))-'0'));
            h = intersect(f,g);
            if length(h)==r-2
                foh = setdiff(f,h); goh = setdiff(g,h);
                alpha = 1*(foh(2)>goh(2) & goh(2)>foh(1)) + 1*(goh(1)>foh(1));
                A_w(2) = A_w(2) + 2^(r+2 + lambda(h,h) + lambda(f,foh) + lambda(g,goh) - alpha);
            end
        end
    end
end
function orbit =  lambda(f,g)
    orbit = 0;
    for i = g
        orbit = orbit + length(setdiff([1:i-1],f));
    end
end
\end{lstlisting}
}

\section{Proof of Lemma \ref{lem:decomp-ltam-gcd}}\label{app:proof-lem:decomp-ltam-gcd}
\begin{IEEEproof}
   Consider an element $P\in\Alow_h\cdot h\cdot \left(\Alow_{f}\cdot \frac{f}{h}+\Alow_g\cdot \frac{g}{h}\right).$ By definition of $\Alow$ and $\Alow_h\cdot h$ we have $P\in \Alow\cdot h\cdot \left(\Alow\cdot \frac{f}{h}+\Alow\cdot \frac{g}{h}\right).$
    Conversely, choose an element $P\in \Alow\cdot h\cdot \left(\Alow\cdot \frac{f}{h}+\Alow\cdot \frac{g}{h}\right)$. We thus have 
    \begin{multline*}
        P=\prod\limits_{i\in\ind(h)}(x_{i}+\sum\limits_{j<i,j\not\in\ind(h)}b_{i,j}x_j+\varepsilon_i)\times\\\prod\limits_{i\in\ind(\frac{f}{h})}(x_{i}+\sum\limits_{j<i,j\not\in\ind(\frac{f}{h})}b_{i,j}x_j+\varepsilon_i)\\
        +\prod\limits_{i\in\ind(h)}(x_{i}+\sum\limits_{j<i,j\not\in\ind(h)}b_{i,j}x_j+\varepsilon_i)\times\\\prod\limits_{i\in\ind(\frac{g}{h})}(x_{i}+\sum\limits_{j<i,j\not\in\ind(\frac{g}{h})}b_{i,j}x_j+\varepsilon_i)
        \end{multline*}
        \begin{multline*}
            P=\prod\limits_{i\in\ind(h)}(x_{i}+\sum\limits_{j<i,j\not\in\ind(h)}b_{i,j}x_j+\varepsilon_i)\times\\\prod\limits_{i\in\ind(\frac{f}{h})}(x_{i}+\sum\limits_{j<i,j\not\in\ind(f)}b_{i,j}^{*}x_j+\varepsilon_i^{*})\\
        +\prod\limits_{i\in\ind(h)}(x_{i}+\sum\limits_{j<i,j\not\in\ind(h)}b_{i,j}x_j+\varepsilon_i)\times\\\prod\limits_{i\in\ind(\frac{g}{h})}(x_{i}+\sum\limits_{j<i,j\not\in\ind(g)}b_{i,j}^{*}x_j+\varepsilon_i^{*})
    \end{multline*}
    where $b_{i,j}^{*}$ are obtained as in the proof of Proposition 3.7.3 from \cite{dragoi17thesis} (see Lemma \ref{lem:prod-lin-form-distinct-var}).
\end{IEEEproof}

\section{Proof Of Theorem \ref{thm:equality-orbits}}\label{app:proof-thm:equality-orbits}

\begin{IEEEproof}
    By Theorem \ref{thm:Kasami-Tokura}, any codeword $\ev(P)$ with $\w(\ev(P))=1.5\wm$ is up to an affine transformation $P=y_1\dots y_{r-2}(y_{r-1}y_{r}+y_{r+1}y_{r+2}).$ 
    
    The direct implication is a consequence of the definition of $\Alow$ and Theorem \ref{thm:Kasami-Tokura}. 
    
    For the converse implication consider a codeword $\ev(P)$ with $P=y_1\dots y_{r-2}(y_{r-1}y_{r}+\dots+y_{r+2\mu-3}y_{r+2\mu-2}).$ The maximum variables in $[y_1\dots y_{r-2}y_{r-1}y_{r}],\dots, [y_1\dots y_{r-2}y_{r+2\mu-3}y_{r+2\mu-2}]$ are all distinct (see Lemma \ref{lem:prod-lin-form-distinct-var}). Hence, we can construct $h=x_{i_1}\dots x_{i_{r-2}}$ and $f/h=x_{i_{r-1}}x_{i_{r}},\dots, g/h=x_{i_{r+1}}x_{i_{r+2}}.$ Notice that while $i_1,\dots i_{r-2}$ are all pairwise distinct, the indices $i_{r-1},i_{r},i_{r+2}$ do not have to be pairwise distinct. Let us demonstrate that there is a affine transformation such that these indices are also pairwise distinct. Suppose $f/h$ and $g/h$ share in common at least 1 maximum variable. Formally, we can write 
    $P_1\in \Alow_{f}\cdot f/h$ as $P_1= (x_{i_1}+v_{i_1})(x_{i_2}+v_{i_2})$ and $P_2 \in \Alow_{g}\cdot g/h$ as $P_2= (x_{j_1}+z_{j_1})(x_{j_2}+z_{j_2})$ where $v_{i_1},v_{i_2},z_{j_1},z_{j_2}$ are linear function with maximum variables strictly smaller that the index, and there is at least one index $i_l$ equal to one index $j_l.$ Also, we have $i_1<i_2$ and $j_1<j_2.$ Then we have the following cases
    \begin{itemize}
        \item $i_1=j_1$ and $i_2>j_2.$ Then 
        \begin{align*}
            P_1+P_2&=(x_{i_1}+v_{i_1})(x_{i_2}+v_{i_2})+(x_{i_1}+z_{i_1})(x_{j_2}+z_{j_2})\\
            &=(x_{i_1}+v_{i_1})(x_{i_2}+v_{i_2}+x_{j_2}+z_{j_2})\\
            &+(x_{j_2}+z_{j_2})(v_{i_1}+z_{i_1})\\
            &\in \Alow_{f}\cdot f/h+\Alow_{g^{*}}\cdot g^{*}/h,            
        \end{align*}
        with $g^{*}/h=x_{j_1^*}x_{j_2}$ where $j_1^{*}=\max\{l \mid x_l \in v_{i_1}+z_{i_1}\}.$ Since $j_1^{*}<i_1$ this implies that all four indices $i_1,i_2,j_1^{*},j_2$ are distinct. Also, $j_1^{*}$ exists since $v_{i_1}\neq z_{i_1}$ and $v_{i_1}\neq 1+z_{i_1}$. If $v_{i_1}= z_{i_1}$ then we would have that $P_1+P_2$ is a product of two linear forms and hence $\ev(P)$ is a minimum weight codeword, which contradicts our hypothesis. If $v_{i_1}= 1+z_{i_1}$ then the weight of $P_1+P_2$ would be 2 times the weight of $P_1$, and this would imply that the weight of $P$ is strictly bigger than $1.5\wm$ which contradicts the hypothesis.
        \item $i_1=j_2.$ Then 
        \begin{align*}
            P_1+P_2&=(x_{i_1}+v_{i_1})(x_{i_2}+v_{i_2})+(x_{j_1}+z_{j_1})(x_{i_1}+z_{i_1})\\
            &=(x_{i_1}+v_{i_1})(x_{i_2}+v_{i_2}+x_{j_1}+z_{j_1})\\
            &+(x_{j_1}+z_{j_1})(v_{i_1}+z_{i_1})\\
            &\in \Alow_{f}\cdot f/h+\Alow_{g^{*}}\cdot g^{*}/h,            
        \end{align*}
        with $g^{*}/h=x_{j_1}x_{j_2^{*}}$ where $j_2^{*}=\max\{l \mid x_l \in v_{i_1}+z_{i_1}\}.$ Here, we might have $j_2^{*}=j_1$ but in that case since $(x_{j_1}+z_{j_1})(v_{i_1}+z_{i_1})$ is a product of linear forms, by Lemma \ref{lem:prod-lin-form-distinct-var} it can be rewritten such that maximum variables are distinct, fact that ends the proof for this case. Also, as in the previous case $j_2^{*}$ exists.
        \item $i_1=j_1$ and $i_2=j_2.$ Then 
        \begin{align*}
            P_1+P_2&=(x_{i_1}+v_{i_1})(x_{i_2}+v_{i_2})+(x_{i_1}+z_{i_1})(x_{i_2}+z_{i_2})
        \end{align*}
        Since $v_{i_1}\neq z_{i_1}$ (or equivalently $\ev(P)$ is not a minimum weight codeword) and $v_{i_2}\neq z_{i_2}$ we have 
        \begin{align*}
            P_1+P_2&=(x_{i_2}+z_{i_2})(z_{i_1}+v_{i_1})+(z_{i_2}+v_{i_2})(x_{i_1}+v_{i_1}).
        \end{align*}
            The two terms in the sum have maximum variables $x_{i_2}x_{j_1^{*}}$ and $x_{j_2^{*}}x_{i_1}$ with $j_1^{*}=\max\{l \mid x_l \in v_{i_1}+z_{i_1}\}$ and $j_2^{*}=\max\{l \mid x_l \in v_{i_2}+z_{i_2}\}.$ If all four indices are distinct the proof is finished. If not, go to the previous items. 
    \end{itemize}
    
\end{IEEEproof}

\section{Proof of Proposition \ref{pr:degree-two-minkowski}}\label{app:proof-pr:degree-two-minkowski}

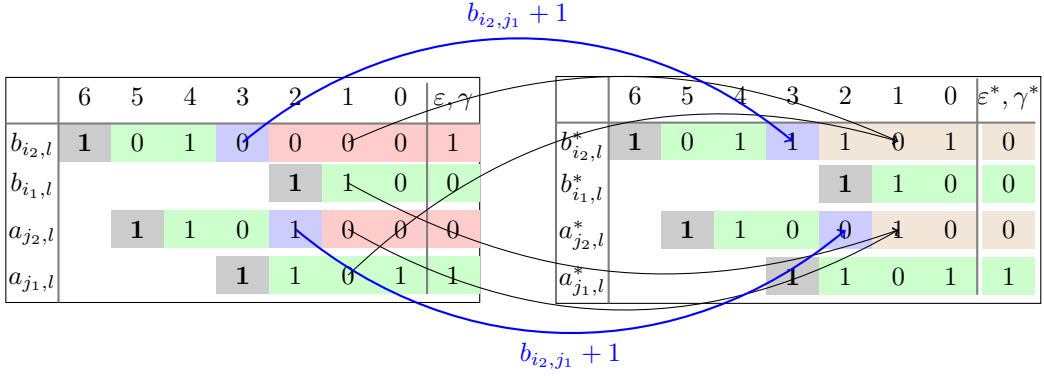
\begin{figure*}[!ht]
\centering
\begin{tikzpicture}
\matrix[draw,inner sep=0pt] (S) [matrix of math nodes,nodes={outer sep=0pt,minimum width=7mm,minimum height=5mm}]
{~&6&5&4&3&2&1&0& \varepsilon,\gamma\\
b_{i_2,l}&|[fill=black!20]|\bm{1}&|[fill=green!20]|0&|[fill=green!20]|1&|[fill=blue!20]|0&|[fill=red!20]|0&|[fill=red!20]|0&|[fill=red!20]|0&|[fill=red!20]|1\\
b_{i_1,l}&~&~&~&~&|[fill=black!20]|\bm{1}&|[fill=green!20]|1&|[fill=green!20]|0&|[fill=green!20]|0\\
a_{j_2,l}&~&|[fill=black!20]|\bm{1}&|[fill=green!20]|1&|[fill=green!20]|0&|[fill=blue!20]|1&|[fill=red!20]|0&|[fill=red!20]|0&|[fill=red!20]|0\\
a_{j_1,l}&~&~&~&|[fill=black!20]|\bm{1}&|[fill=green!20]|1&|[fill=green!20]|0&|[fill=green!20]|1&|[fill=green!20]|1\\
};

\matrix[draw,inner sep=0pt] (T) [matrix of math nodes,nodes={outer sep=0pt,minimum width=7mm,minimum height=5mm},right=of S]
{~&6&5&4&3&2&1&0& \varepsilon^{*},\gamma^{*}\\
b_{i_2,l}^{*}&|[fill=black!20]|\bm{1}&|[fill=green!20]|0&|[fill=green!20]|1&|[fill=blue!20]|1&|[fill=brown!20]|1&|[fill=brown!20]|0&|[fill=brown!20]|1&|[fill=brown!20]|0\\
b_{i_1,l}^{*}&~&~&~&~&|[fill=black!20]|\bm{1}&|[fill=green!20]|1&|[fill=green!20]|0&|[fill=green!20]|0\\
a_{j_2,l}^{*}&~&|[fill=black!20]|\bm{1}&|[fill=green!20]|1&|[fill=green!20]|0&|[fill=blue!20]|0&|[fill=brown!20]|1&|[fill=brown!20]|0&|[fill=brown!20]|0\\
a_{j_1,l}^{*}&~&~&~&|[fill=black!20]|\bm{1}&|[fill=green!20]|1&|[fill=green!20]|0&|[fill=green!20]|1&|[fill=green!20]|1\\
};

\draw[thick,gray] (S-1-2.north west) -- (S-5-2.south west);
\draw[thick,gray] (S-1-1.south west) -- (S-2-9.north east);
\draw[thick,gray] (S-1-9.north west) -- (S-5-9.south west);

\draw[thick,gray] (T-1-2.north west) -- (T-5-2.south west);
\draw[thick,gray] (T-1-1.south west) -- (T-2-9.north east);
\draw[thick,gray] (T-1-9.north west) -- (T-5-8.south east);

\path [->] (S-2-7.center) edge [bend left=24] node{} (T-2-7.center);
\path [->] (S-5-7.center) edge [bend left=30] node{} (T-2-7.center);

\path [->] (S-3-7.center) edge [bend right=24] node{} (T-4-7.center);
\path [->] (S-4-7.center) edge [bend right=30] node{} (T-4-7.center);

\path [->,blue,thick] (S-2-5.center) edge [bend left=40] node[above]{$b_{i_2,j_1}+1$} (T-2-5.center);
\path [->,blue,thick] (S-4-6.center) edge [bend right=40] node[below]{$b_{i_2,j_1}+1$} (T-4-6.center);

\end{tikzpicture}
\caption{Application of Proposition \ref{pr:degree-two-minkowski} in the case $i_2>j_2>j_1>i_1$ with $f=x_6x_2,g=x_5x_3.$ For $P=(x_6+x_4+1)(x_2+x_1)\in \Alow\cdot f$ and $Q=(x_5+x_4+x_2)(x_3+x_2+x_0+1)\in\Alow\cdot g$ we can create a non-trivial distinct pair $P^{*},Q^{*}$ with $P^{*}=(x_6+x_4+x_3+x_2+x_0)(x_2+x_1)$ and $Q^{*}=(x_5+x_4+x_1)(x_3+x_2+x_0+1)$. The meaning of the colors is: gray cells are for indices of variables in the support, green cells are for $b_{i,j}^{*}=b_{i,j}$ or $a_{i,j}^{*}=a_{i,j}$, blue cells are denoting $b_{i,j}^{*}=1+b_{i,j}$ or $a_{i,j}^{*}=1+a_{i,j}$, and brown cells are obtained by the combination of red cells and their green counterparts; that is, $b_{i_2,l}^{*}=b_{i_2,l}+a_{j_1,l}$ and $a_{j_2,l}^{*}=a_{j_2,l}+b_{i_1,l}$.}\label{fig:ex-pr2Minkowski} 
\end{figure*}

Let $f=x_{i_1}x_{i_2}$ and $g=x_{j_1}x_{j_2}$ with $\gcd(f,g)=1$ and $i_2>j_2$ and $P=(\bB,\bve)\cdot f,P=(\bB^{*},\bve^{*})\cdot f, Q=(\bA,\bvg)\cdot g,Q^{*}=(\bA^{*},\bvg^{*})\cdot g.$ By definition of $\Alow$ and of a collision we have 
    \begin{multline}
        (x_{i_2}+\sum\limits_{l<i_2,l\neq i_1}b_{i_2,l}x_l+\varepsilon_{i_2})(x_{i_1}+\sum\limits_{l<i_1}b_{i_1,l}x_l+\varepsilon_{i_1})\\-
        (x_{i_2}+\sum\limits_{l<i_2,l\neq i_1}b_{i_2,l}^{*}x_l+\varepsilon_{i_2}^{*})(x_{i_1}+\sum\limits_{l<i_1}b_{i_1,l}^{*}x_l+\varepsilon_{i_1}^{*})\\=
        (x_{j_2}+\sum\limits_{l<j_2,l\neq j_1}a_{j_2,l}x_l+\gamma_{j_2})(x_{j_1}+\sum\limits_{l<j_1}a_{j_1,l}x_l+\gamma_{j_1})\\-
        (x_{j_2}+\sum\limits_{l<j_2,l\neq j_1}a_{j_2,l}^{*}x_l+\gamma_{j_2}^{*})(x_{j_1}+\sum\limits_{l<j_1}a_{j_1,l}^{*}x_l+\gamma_{j_1}^{*})
    \end{multline}

Since $x_{i_2}$ is the maximum variable, extracting the coefficient of $x_{i_2}$ gives 
 \begin{equation*}
     x_{i_1}+\sum\limits_{l<i_1}b_{i_1,l}x_l+\varepsilon_{i_1}=x_{i_1}+\sum\limits_{l<i_1}b_{i_1,l}^{*}x_l+\varepsilon_{i_1}^{*}.
 \end{equation*}
 This implies 
 \begin{multline}
     (\sum\limits_{l<i_2,l\neq i_1}(b_{i_2,l}-b_{i_2,l}^{*})x_l+\varepsilon_{i_2}-\varepsilon_{i_2}^{*})(x_{i_1}+\sum\limits_{l<i_1}b_{i_1,l}x_l+\varepsilon_{i_1})\\=Q-Q^{*}.
 \end{multline}
If we are in the case $i_2>i_1>j_2.$ Then either $x_l$ is the maximum variable (in the first factor of the first term) or $x_{i_1}$ is the maximum variable (in the second factor of the first term). Either ways we deduce $Q-Q^{*}=0$ which implies $\bA=\bA^{*}$ and further no possible collisions. Hence, in this case the Minkowski sum $\Alow\cdot f+\Alow\cdot g$ has maximum cardinality.
 
 If we are in the case $i_2>j_2>i_1.$ If we consider that the next maximum variable is $x_l$ (present in the first term), with $j_2<l<i_2$ we obtain $b_{i_2,l}=b_{i_2,l}^{*}$ for all such indices. This means that we have 
 \begin{multline}
     (\sum\limits_{l\leq {\color{blue}j_2},l\neq i_1}(b_{i_2,l}-b_{i_2,l}^{*})x_l+\varepsilon_{i_2}-\varepsilon_{i_2}^{*})(x_{i_1}+\sum\limits_{l<i_1}b_{i_1,l}x_l+\varepsilon_{i_1})\\=Q-Q^{*}.
 \end{multline}

 If the next maximum variable is $x_{j_2}$ we have two cases, $b_{i_2,j_2}-b_{i_2,j_2}^{*}=0$ and $b_{i_2,j_2}-b_{i_2,j_2}^{*}=1.$ 
 
 {\textbf{A) The case $b_{i_2,j_2}-b_{i_2,j_2}^{*}=0$}.} The last equation becomes
  \begin{multline}
     (\sum\limits_{l{\color{blue}<} j_2,l\neq i_1}(b_{i_2,l}-b_{i_2,l}^{*})x_l+\varepsilon_{i_2}-\varepsilon_{i_2}^{*})(x_{i_1}+\sum\limits_{l<i_1}b_{i_1,l}x_l+\varepsilon_{i_1})\\=Q-Q^{*}.
 \end{multline}

Extracting the coefficient of $x_{j_2}$ gives
 \begin{equation*}
     x_{j_1}+\sum\limits_{l<j_1}a_{j_1,l}x_l+\gamma_{j_1}=x_{j_1}+\sum\limits_{l<j_1}a_{j_1,l}^{*}x_l+\gamma_{j_1}^{*}.
 \end{equation*} which implies
 \begin{multline}\label{eq:20}
     (\sum\limits_{l<j_2,l\neq i_1}(b_{i_2,l}-b_{i_2,l}^{*})x_l+\varepsilon_{i_2}-\varepsilon_{i_2}^{*})(x_{i_1}+\sum\limits_{l<i_1}b_{i_1,l}x_l+\varepsilon_{i_1})\\=(\sum\limits_{l<j_2,l\neq j_1}(a_{j_2,l}-a_{j_2,l}^{*})x_l+\gamma_{j_2}-\gamma_{j_2}^{*})(x_{j_1}+\sum\limits_{l<j_1}a_{j_1,l}x_l+\gamma_{j_1}).
 \end{multline}

In order to have equation \eqref{eq:20} valid we need 
\begin{itemize}
    \item $b_{i_2,l}=b_{i_2,l}^{*}$ for $j_1<l<j_2$ 
    \item $a_{j_2,l}=a_{j_2,l}^{*}$ for $i_1<l<j_2$
    \item $b_{i_2,j_1}-b_{i_2,j_1}^{*}=1$
    \item $a_{j_2,i_1}-a_{j_2,i_1}^{*}=1$
    \item $b_{i_2,l}-b_{i_2,l}^{*}=a_{j_1,l}$ for $l<j_1$
    \item $a_{j_2,l}-a_{j_2,l}^{*}=b_{i_1,l}$ for $l<i_1$
    \item $\varepsilon_{i_2}-\varepsilon_{i_2}^{*}=\gamma_{j_1}$
    \item $\gamma_{j_2}-\gamma_{j_2}^{*}=\varepsilon_{i_1}$

    Since these are all the possible coefficients this case ends here.
\end{itemize} 
{\textbf{B) The case $b_{i_2,j_2}-b_{i_2,j_2}^{*}=1$}.}  We have 
      \begin{multline}
     (x_{j_2}+\sum\limits_{l< j_2,l\neq i_1}(b_{i_2,l}-b_{i_2,l}^{*})x_l+\varepsilon_{i_2}-\varepsilon_{i_2}^{*})(x_{i_1}+\sum\limits_{l<i_1}b_{i_1,l}x_l+\varepsilon_{i_1})\\=Q-Q^{*}.
 \end{multline}
Extracting the coefficient of $x_{j_2}$ gives
 \begin{multline*}
     x_{i_1}+\sum\limits_{l<i_1}b_{i_1,l}x_l+\varepsilon_{i_1}=\sum\limits_{l<j_1}(a_{j_1,l}-a_{j_1,l}^{*})x_l+\gamma_{j_1}-\gamma_{j_1}^{*}.
 \end{multline*}
    This equation can hold only if $i_1<j_1.$ If this is the case then we deduce
\begin{align*}
 a_{j_1,l}-a_{j1,l}^{*}&=0, \;for\;i_1<l<j_1\\
    a_{j_1,i_1}-a_{j1,i_1}^{*}&=1,\\
    a_{j_1,l}-a_{j1,l}^{*}&=b_{i_1,l}, \;for\;l<i_1\\
    \gamma_{j_1}-\gamma_{j_1}^{*}=\varepsilon_{i_1}.
\end{align*}
   Also we deduce 
      \begin{multline}
     (x_{j_2}+\sum\limits_{l< j_2,l\neq i_1}(b_{i_2,l}-b_{i_2,l}^{*})x_l+\varepsilon_{i_2}-\varepsilon_{i_2}^{*})(x_{i_1}+\sum\limits_{l<i_1}b_{i_1,l}x_l+\varepsilon_{i_1})\\=
     (x_{j_2}+\sum\limits_{l<j_2,l\neq j_1}a_{j_2,l}x_l+\gamma_{j_2})(x_{j_1}+\sum\limits_{l\leq i_1}a_{j_1,l}x_l+\gamma_{j_1})\\-
        (x_{j_2}+\sum\limits_{l<j_2,l\neq j_1}a_{j_2,l}^{*}x_l+\gamma_{j_2}^{})(x_{j_1}+\sum\limits_{l\leq i_1}a_{j_1,l}^{}x_l+\gamma_{j_1}^{})\\
        -(x_{j_2}+\sum\limits_{l<j_2,l\neq j_1}a_{j_2,l}^{*}x_l+\gamma_{j_2}^{})(x_{i_1}+\sum\limits_{l<i_1}b_{i_1,l}x_l+\varepsilon_{i_1})
 \end{multline}
 which leads to
      \begin{multline}\label{eq:23}
     (\sum\limits_{l< j_2,l\neq i_1}(b_{i_2,l}-b_{i_2,l}^{*})x_l+\varepsilon_{i_2}-\varepsilon_{i_2}^{*}-\sum\limits_{l<j_2,l\neq j_1}a_{j_2,l}^{*}x_l-\gamma_{j_2}^{})\\\times(x_{i_1}+\sum\limits_{l<i_1}b_{i_1,l}x_l+\varepsilon_{i_1})\\=
     (\sum\limits_{l<j_2,l\neq j_1}(a_{j_2,l}-a_{j_2,l}^{*})x_l+\gamma_{j_2}-\gamma_{j_2}^{*})(x_{j_1}+\sum\limits_{l\leq i_1}a_{j_1,l}x_l+\gamma_{j_1})
 \end{multline}
 
Notice that one can not have $x_l$ with $l>j_1$ as maximum variable in the previous equation, i.e., we need to have $a_{j_2,l}=a_{j_2,l}^{*}$ for $l>i_1$ and $b_{i_2,l}-b_{i_2,l}^{*}-a_{j_2,l}^{*}=0$ for $l>j_1.$ And as in the previous case we can determine a set of conditions under which equation \eqref{eq:23} is valid. 
\begin{itemize}
    \item $b_{i_2,j_1}-b_{i_2,j_1}^{*}=1$
    \item $a_{j_2,i_1}-a_{j_2,i_1}^{*}=1$
    \item $b_{i_2,l}-b_{i_2,l}^{*}=a_{j_1,l}-a_{j_2,l}^{*}$ for $l<j_1$
    \item $a_{j_2,l}-a_{j_2,l}^{*}=b_{i_1,l}$ for $l<i_1$
    \item $\varepsilon_{i_2}-\varepsilon_{i_2}^{*}=\gamma_{j_1}-\gamma_{j_2}$
    \item $\gamma_{j_2}-\gamma_{j_2}^{*}=\varepsilon_{i_1}$
\end{itemize} 
Since, all possible coefficients are present, this case ends here.

Resuming the cases, when $i_2>i_1>j_2$ no collisions, when $i_2>j_2>j_1>i_2$ we have two restrictions, and when $i_2>j_2>i_1>j_1$ we have a single restriction. Each restriction is an equation between two free variables ($b_{i_2,j_2},b_{i_2,j_2}^{*}$) over $\ff_2$, hence, each restriction generates 2 possible solutions, from which we deduce the wanted result. In other words, the variable $b_{i_2,j_2}$ is a free variable in $\bB$ that allows us to count for each element the number of invariants. 

\section{Proof of Lemma \ref{lem:cardinal-product-orbits}}

Let us suppose by absurd that there are polynomials $H,H^{*}\in\Alow_h\cdot h$ and $P,P^{*}\in \Alow_{f}\cdot \frac{f}{h}+\Alow_g\cdot \frac{g}{h}$ with $H\neq H^{*}$ and $P\neq P^{*}$ s.t. $HP=H^{*}P^{*}.$ Since $h$ is a product of variables that are not present in $P$ or $P^{*}$ extracting the coefficient of $h$ from $HP$ and $H^{*}P^{*}$ implies $P=P^{*}.$ But this would contradict our hypothesis, and ends the proof.

\section{Proof of Proposition \ref{pr:no-colision-1.5dorbits}}\label{app:proof-pr:no-colision-1.5dorbits}

\begin{IEEEproof}
Suppose by absurd that there is a common polynomial $P$ between the two sets. Since $P$ belongs to $\Alow_h\cdot h\cdot(\Alow_{f} \cdot \frac{f}{h}+\Alow_{g}\cdot \frac{g}{h})$ this implies that the maximum monomials in $P$ are $f+g$ (by Definition of the $\Alow$). Also since $P$ belongs to $\Alow_h^{*}\cdot h^{*}\cdot(\Alow_{f^{*}} \cdot \frac{f^{*}}{h^{*}}+\Alow_{g^{*}}\cdot \frac{g^{*}}{h^{*}})$ this implies that the maximum monomials in $P$ are $f^{*}+g^{*}.$ Hence, $f+g=f^{*}+g^{*}.$ Both sums can not be equal to zero since $f\neq g$ and $f^{*}\neq g^{*}.$ Hence, we need to have either $(f,g)=(f^{*},g^{*})$ or $(f,g)=(g^{*},f^{*})$ which ends the proof.    
\end{IEEEproof}

\section{Proof of Theorem \ref{thm:formula_15w}}\label{app:proof-thm:formula_15w}

\begin{IEEEproof}
 First, the cardinality of $W_{1.5\wm}$ can be computed as the sum of $|\Alow_h\cdot h|\times|(\Alow_{f}\cdot x_{i_1}x_{i_2}+\Alow_{g} \cdot x_{j_1}x_{j_2})|$ for all possible $h$ with $\deg(h)=r-2$ and $f=hx_{i_1}x_{i_2},g=hx_{j_1}x_{j_2}$ \\
 Second, recall that $|\Alow_h\cdot g|=2^{\deg(g)+|\lambda_h(g)|}$ and $|\Alow_h\cdot h|=2^{\deg(h)+|\lambda_h|}.$ \\
    Combined with Propostion \ref{pr:degree-two-minkowski} we have that $|(\Alow_{f}\cdot x_{i_1}x_{i_2}+\Alow_{g} \cdot x_{j_1}x_{j_2})|=|(\Alow_{f}\cdot x_{i_1}x_{i_2}|\times|\Alow_{g} \cdot x_{j_1}x_{j_2})|\times 2^{\alpha_{\frac{f}{h},\frac{g}{h}}}$. 
    Hence, each orbit has cardinality $2^{r-2+|\lambda_h|}\times 2^{2+|\lambda_{f}(\frac{f}{h})|+2+|\lambda_{g}(\frac{g}{h})|-\alpha_{\frac{f}{h},\frac{g}{h}}}$, which ends the proof. 
\end{IEEEproof}
\end{document}